\documentclass[fleqn,usenatbib]{mnras}

\usepackage{newtxtext,newtxmath}

\usepackage[T1]{fontenc}

\DeclareRobustCommand{\VAN}[3]{#2}
\let\VANthebibliography\thebibliography
\def\thebibliography{\DeclareRobustCommand{\VAN}[3]{##3}\VANthebibliography}

\usepackage{graphicx}	
\usepackage{amsmath}	
\usepackage{stfloats}
\usepackage{multirow}
\usepackage[normalem]{ulem}
\usepackage{hyperref}

\title[Finding AGN with \textit{JWST} MIRI]{Finding dusty AGNs from the \textit{JWST} CEERS survey with mid-infrared photometry}

\author[Tom C.-C. Chien et al.]{
Tom C.-C. Chien$^{1}$,\thanks{E-mail: minoodos90214@gmail.com}
Chih-Teng Ling$^{2}$,
Tomotsugu Goto$^{1,2}$,
Cossas K.-W. Wu$^{2}$,
Seong Jin Kim$^{2}$,
\newauthor
Tetsuya Hashimoto$^{3}$,
Yu-Wei Lin$^{1}$,
Ece Kilerci$^{4}$,
Simon C.-C. Ho$^{5,6,7,8}$,
Po-Ya Wang$^{1}$, and
\newauthor
Bjorn Jasper R. Raquel$^{3,9}$
\\
$^{1}$Department of Physics, National Tsing Hua University, 101, Section 2. Kuang-Fu Road, Hsinchu, 30013, Taiwan (R.O.C.)\\
$^{2}$Institute of Astronomy, National Tsing Hua University, 101, Section 2. Kuang-Fu Road, Hsinchu, 30013, Taiwan (R.O.C.)\\
$^{3}$Department of Physics, National Chung Hsing University, 145, Xingda Road, Taichung, 40227, Taiwan (R.O.C.)\\
$^{4}$Sabanc{\i} University, Faculty of Engineering and Natural Sciences, 34956, Istanbul, Turkey\\
$^{5}$Research School of Astronomy and Astrophysics, The Australian National University, Canberra, ACT 2611, Australia\\
$^{6}$Centre for Astrophysics and Supercomputing, Swinburne University of Technology, P.O. Box 218, Hawthorn, VIC 3122, Australia\\
$^{7}$OzGrav: The Australian Research Council Centre of Excellence for Gravitational Wave Discovery, Hawthorn, VIC 3122, Australia\\
$^{8}$ASTRO3D: ARC Centre of Excellence for All-sky Astrophysics in 3D, ACT 2611, Australia\\
$^{9}$National Institute of Physics, College of Science, University of the Philippines, Diliman, Quezon City, 1101 Metro Manila, Philippines
}

\date{Accepted 2024 June 19. Received 2024 June 17; in original form 2023 September 29}

\pubyear{2024}

\begin{document}
\label{firstpage}
\pagerange{\pageref{firstpage}--\pageref{lastpage}}
\maketitle

\begin{abstract}
The nature of the interaction between active galactic nuclei (AGNs) and their host galaxies remains an unsolved question. Therefore, conducting an AGN census is valuable to AGN research. Nevertheless, a significant fraction of AGNs are obscured by their environment, which blocks UV and optical emissions due to the dusty torus surrounding the central supermassive black hole (SMBH). To overcome this challenge, mid-infrared (IR) surveys have emerged as a valuable tool for identifying obscured AGNs, as the obscured light is re-emitted in this range. With its high sensitivity, the James Webb Space Telescope (\textit{JWST}) uncovered more fainter objects than previous telescopes. By applying the SED fitting, this work investigates AGN candidates in \textit{JWST} Cosmic Evolution Early Release Science (CEERS) fields. We identified 42 candidates, 30 of them are classified as composites ($0.2\leq f_{\rm AGN, IR}< 0.5$), and 12 of them are AGNs ($f_{\rm AGN, IR}\geq 0.5$). We report the AGN luminosity contributions and AGN number fractions as a function of redshift and total infrared luminosity, showing that previously reported increasing relations are not apparent in our sample due to the sample size. We also extend the previous results on ultra-luminous infrared galaxies (ULIRGs, $L_{\rm TIR}\geq 10^{12} L_{\odot}$) to less luminous AGNs, highlighting the power of \textit{JWST}.
\end{abstract}




\section{Introduction}\label{intro}
Active galactic nuclei (AGNs) are one of the most intriguing and challenging phenomena in modern astronomy. Supermassive black holes (SMBHs) residing in AGNs trigger many spectacular phenomena \citep[e.g., relativistic jets,][]{Blandford2019}{}{}, brightening AGNs and influencing the evolution of their host galaxies. Therefore, studying AGNs provides a valuable way to understand the co-evolution of SMBHs and their host galaxies, which is crucial for understanding how galaxies evolve. There are several features in AGNs according to the AGN unification model \citep[e.g.,][]{Netzer2015}, for instance, accretion disc around centre SMBH, broad line and narrow line region, and dusty torus which is more outer than accretion disc. One of the important features is the hot accretion disk around SMBH, which emits UV/optical radiation. Nevertheless, due to the obscuration by the surrounding dust, the majority of AGNs are missed in UV and optical bands, and even hard X-ray emission could be obscured in a Compton-thick AGN \citep[e.g.,][]{Comastri2004}. 
Due to the nature of obscuration in these bands, it is usually difficult to find dusty AGNs directly. To prevent missing obscured objects, a mid-infrared (mid-IR) observation survey becomes significant when researching obscured AGNs \citep[e.g.,][]{Chang2017, Wang2020}{}{}. Additionally, several studies have reported a strong correlation between mid-infrared (mid-IR) luminosity and X-ray luminosity \citep[e.g.,][]{Auge2023}{}{}, showing the significance and effectiveness of mid-IR observations in probing AGNs.

The absorbed UV/optical light can be re-emitted in mid-IR wavelengths by the surrounding dusty environments. Such environments make the obscured AGNs extinction-less in mid-IR wavelengths. Besides, the majority of the high redshift (high-z) galaxies are identified as obscured objects \citep[e.g.,][]{Treister2010}. Their existence is substantial for probing the history of SMBHs, which further connects to the black hole accretion history \citep[BHAH, e.g.,][]{Yang2023}{}{}. Therefore, understanding the properties of AGNs allows the study to investigate the history of centre SMBHs and explore the co-evolution between SMBHs and their host galaxies. Nevertheless, it is worth noting that the emission of polycyclic aromatic hydrocarbon (PAH) in mid-IR wavelengths produced by star-forming galaxies (SFGs) is an issue in identifying AGNs \citep{Feltre2013, Kim2019}. It substantially impacts the identification by mimicking AGN emissions in mid-IR wavelength. To avoid such sample contamination, spectral energy distribution (SED) is a powerful tool that allows us to significantly divide SFGs and AGNs. For instance, \cite{Yang2023} applied Code Investigating GALaxy Emission v2022.1 \citep[\textsc{CIGALE};][]{Boquien2019}{}{}, a SED fitting code, to analysed and classified observed sources into AGNs, composites, and SFGs. They also utilised the physical properties estimated from \textsc{CIGALE} to depict the BHAH.

With mid-IR observation, it is possible to unearth obscured AGNs due to the strong emission from their dusty torus. Recently, the state-of-the-art IR telescope, James Webb Space Telescope (\textit{JWST}) provided an unprecedented sensitivity to observe fainter galaxies. The Mid-Infrared Instrument \citep[MIRI,][]{Rieke2015, Rigby2023}{}{} is the main instrument to study obscured AGNs. \cite{Lyu2024} utilised MIRI in Systematic Mid-infrared Instrument Legacy Extragalactic Survey \citep[SMILES,][]{Rieke2017}{}{} and identified a surprisingly large number of new obscured AGNs ($\sim$80\% among their 217 candidates). This highlights the potential of \textit{JWST} as the premier instrument for investigating obscured AGNs. Back to MIRI, it has a continuous coverage in the mid-IR region (from $5.6\mu m$ to $25.5\mu m$, 9 bands), which is much more sensitive than \textit{AKARI} and the \textit{Spitzer} up to 100 times they are \citep{Wu2023, Ling2022}. 

Previous works reported an increasing AGN contribution trend with redshift \citep[e.g.,][]{Wang2020, Chiang2019}. \cite{Wang2020} performed an extinction-less census of AGNs in the North Ecliptic Pole \citep[NEP,][]{Matsuhara2006} with the \textit{AKARI} space telescope \citep[][]{Murakami2007}. The AGN contribution ($f_{\rm AGN, IR}$, defined in equation \ref{eq:eq1}) and AGN number fraction ($f_{num}$, defined in equation \ref{eq:eq2}) are investigated in \cite{Wang2020} by utilising \textsc{CIGALE} for SED fitting, where an increasing trend with redshift for both quantities is reported. However, it is unclear if the trend is universal for AGNs in different total IR luminosity ranges. \cite{Chiang2019} provided an extinction-less census in the same survey as well, however, the SED fitting was performed by \textsc{LePhare} \citep{Ilbert2006, Arnouts1999}. The results were opposite to \cite{Wang2020}. In this work, our goal is to examine the trend from \cite{Wang2020} still holds at higher redshift or lower luminosity with our AGN sample selected from the Cosmic Evolution Early Release Science \citep[CEERS,][]{Finkelstein2017, yang2023ceers} survey. \textit{JWST} provides more sensitive detection to observe fainter sources \citep{Wu2023, Ling2022}. We expect fainter sources that are invisible for previous space telescopes can extend \cite{Wang2020} and \cite{Chiang2019} researches to the faint end ($L_{\rm TIR}\sim10^{10}L_{\odot}$ or even fainter).
 
The structure of this paper is the following: We introduce the data processes, SED fittings, and sample selections in \S \ref{Data_analysis}; In \S \ref{RandD}, we discuss the properties of our candidates based on our analysis results. The conclusion part summarises our main results, given in \S \ref{conclusion}. We assume a cosmology based on $\Lambda$ cold dark matter ($\Lambda$CDM) cosmology with $H_{0}=0.667$, $\Omega_{m}=0.310$, $\Omega_{b}=0.0490$, and $\Omega_{\Lambda}=0.689$ \citep{Planck2020A&A...641A...6P}.

\section{Data and analysis}\label{Data_analysis}

\subsection{MIR observations in CEERS and merged catalogue with CANDELS-EGS}
\textit{JWST} Cosmic Evolution Early Release Science \citep[CEERS,][]{Finkelstein2017, yang2023ceers} survey provides a continuous observation from near-infrared (NIR) to mid-infrared (MIR) wavelength by \textit{JWST} Near Infrared Camera (NIRcam), Mid-Infrared Instrument \citep[MIRI,][]{Rieke2015, Rigby2023}, and Near-Infrared Spectrograph (NIRSpec). 

In this work, we focus on 4 MIRI pointings in the Extended Groth Strip (EGS) legacy field with 6 continuous broad-band filters (F770W, F1000W, F1280W, F1500W, F1800W, and F2100W). Two pointings have all 6 bands (observation ID: o001\_t021 and o002\_t022), while the other two pointings have only 4 bands, missing F770W and F2100W (observation ID: o012\_t026 and o015\_t028). We use photometry reported in \cite{Wu2023} which conducts a series of processes including 80\% completeness measurements and comparisons with previous model predictions \citep[e.g.,][]{Gruppioni2011, Cowley2018}. Besides, given that STScI has released the new MIRI flux calibration (see \href{https://www.stsci.edu/contents/news/jwst/2023/updates-to-the-miri-imager-flux-calibration-reference-files}{https://www.stsci.edu/contents/news/jwst/2023/updates-to-the-miri-imager-flux-calibration-reference-files}), we have followed the manuscripts to update the photometry as well. The related updates are 15.0\%, 4.1\%, 0.8\%, 1.2\%, 3.3\%, and 3.3\% higher than previous for F770W, F1000W, F1280W, F1500W, F1800W, and F2100W, respectively. \cite{Wu2023}'s 80\% completeness flux in CEERS survey are also scaled for F770W ($0.28 \mu Jy$), F1000W ($0.65 \mu Jy$), F1280W ($1.26 \mu Jy$), F1500W ($2.0 \mu Jy$), F1800W ($5.0 \mu Jy$), and F2100W ($13.4 \mu Jy$), respectively. More technical details are described in \cite{Wu2023}.

To study the physical properties of AGNs, it is insufficient to utilise only MIR photometry. Therefore, a photometric catalogue covering multiple wavelengths is necessary. CANDELS-EGS Multi-wavelength catalogue (\citealt{Stefanon2017}) constructed from the Cosmic Assembly Near-IR Deep Extra-galactic Legacy Survey \citep[CANDELS:][]{Grogin2011, Koekemoer2011} offers a broad observation range from near-UV to MIR in the EGS field. By matching our sources with CANDELS-EGS, we obtain a 21 multi-band catalogue with 15 bands from CANDELS-EGS and 6 bands from CEERS MIRI pointings (\citealt{Ling2024}, see Table \ref{tab:multibands} for details). Our catalogue includes 573 matched sources, each with at least one MIRI detection.
\begin{table*}
	\centering
	\begin{tabular}{cccc}
		\hline
		Telescope & Instrument & Bands & Reference\\
		\hline
		CFHT & MegaCam & $u^{*}$, $g'$, $r'$, $i'$, $z'$ & \cite{Gwyn2012}\\
	    CFHT & WIRCam & J, H, Ks & \cite{Bielby2012}\\
            \hline
		\textit{HST} & ACS & F606W, F814W & \cite{Koekemoer2011}\\
            \textit{HST} & WFC3 & F125W, F160W & \cite{Koekemoer2011}\\
            \textit{HST} & WFC3 & F140W & \cite{Skelton2014}\\
            \hline
            \textit{Spitzer} & IRAC & 3.6$\mu m$, 4.5$\mu m$ & \cite{Ashby2015}\\
            \hline
            \textit{JWST} & MIRI & F770W, F1000W, F1280W, F1500W, F1800W, F2100W & \cite{Ling2024, Wu2023}\\
		\hline
	\end{tabular}
        \caption{The details of bands used in merged multi-wavelength catalogue.}
        \label{tab:multibands}
\end{table*}

\subsection{CIGALE SED fittings}\label{CIGALE}
Code Investigating GALaxy Emission v2022.1 \citep[\textsc{CIGALE};][]{Boquien2019}{}{} is a Python-based code for fitting spectral energy distributions (SEDs) and can estimate the physical properties of galaxies with observations from far-UV to radio emission. It provides various modules such as star-forming history, dust attenuation, dust model, and AGN model to fit observations. In this work, we mainly follow the configuration from \cite{Yang2023}. We slightly change the number of AGN fractions, the dust attenuation model, and the given redshift. In Table \ref{tab:SED_config}, we list the details for the full parameters we adopt.

Note that in \textsc{CIGALE}, AGN contribution is called AGN fraction. Other literature might treat AGN fraction as AGN number fraction \citep[e.g.,][]{Gu2018}{}{}, which is another main discussion in this work. To avoid confusion, here, we use the term AGN contribution to replace the AGN fraction.
\begin{figure*}
    \includegraphics[width=0.45\textwidth]{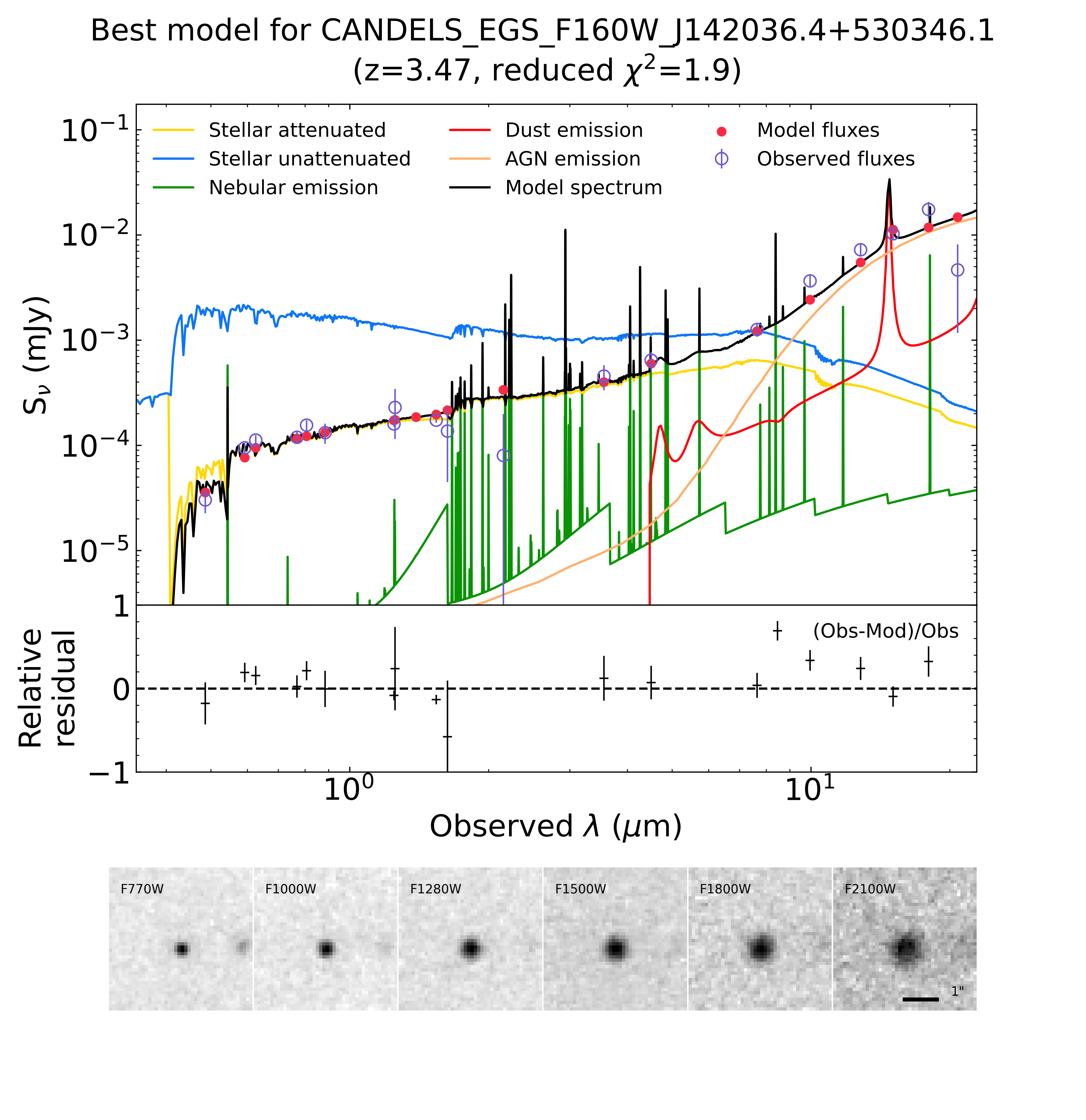}
    \includegraphics[width=0.45\textwidth]{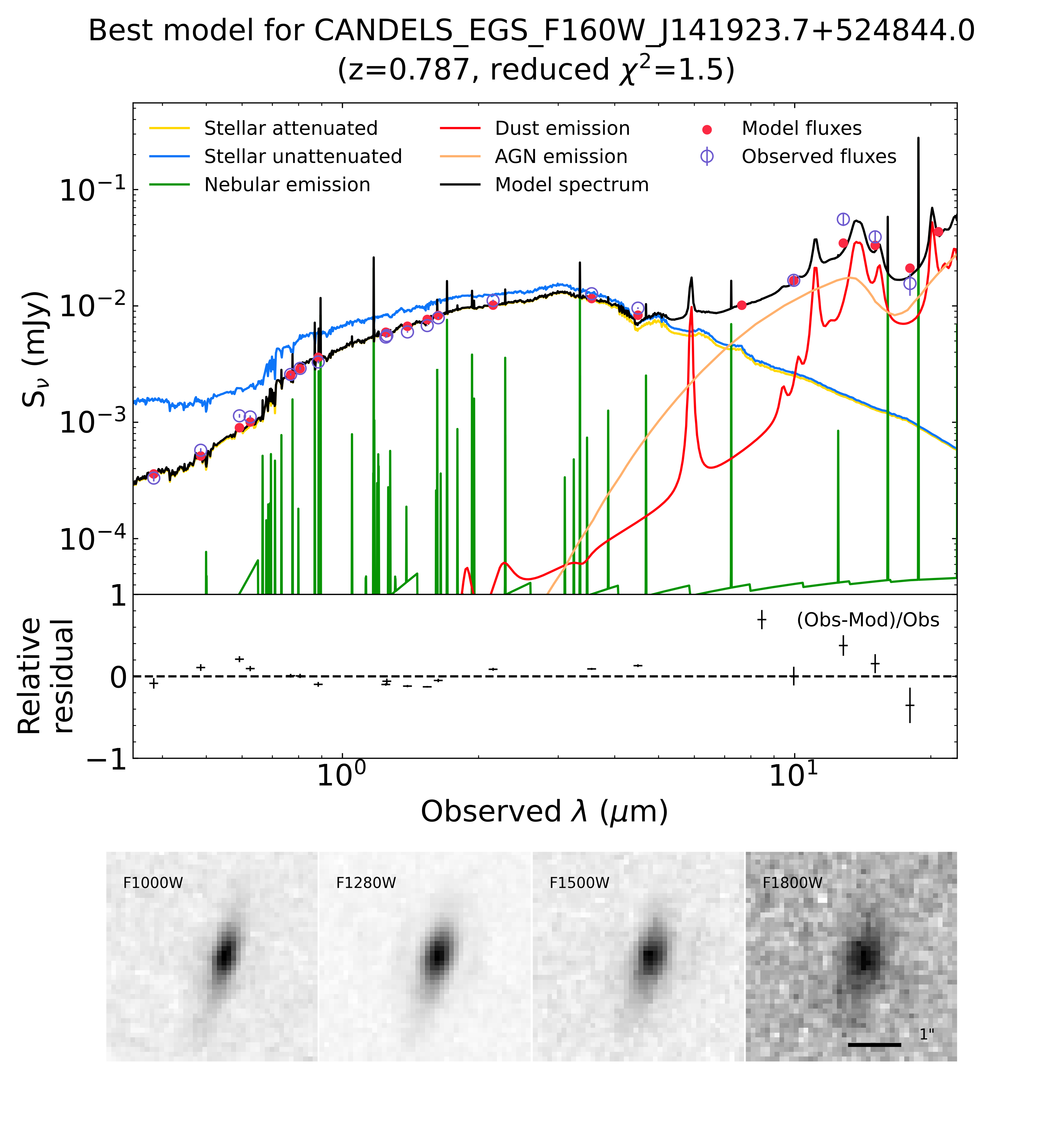}
    \caption{Two SEDs for AGN ($f_{\rm AGN, IR}\geq 0.5$, left panel) and composite ($0.2\leq f_{\rm AGN, IR}< 0.5$, right panel) galaxy SED fitting examples from \textsc{CIGALE} and their cutout. The black line is the fitting curve; The red line is the dust emission, and the orange line is the AGN emission. The AGN contributions are $f_{\rm AGN, IR}=0.66\pm 0.21$ and $f_{\rm AGN, IR}=0.30\pm 0.20$, respectively.}
    \label{fig:SEDcut1}
\end{figure*}

AGN contribution ($f_{\rm AGN, IR}$) is the ratio between AGN luminosity ($L_{AGN}$) and total infrared luminosity, describing how much the AGN emission contributes to the total emission. The definition of it is described in equation~\ref{eq:eq1}:
\begin{equation}
    f_{\rm AGN, IR} = \dfrac{L_{\rm AGN}}{L_{\rm TIR}}.
    \label{eq:eq1}
\end{equation}
where IR is defined in $8-1000\mu m$ by \citealt{Kennicutt1998}. The default wavelength range to define the AGN contribution in \textsc{CIGALE} is $8\,\mu m-1000\,\mu m$. Note that the total IR luminosity estimated by \textsc{CIGALE} might have uncertainty due to a lack of far-infrared (FIR) observations in this work. The relative problems will be discussed in \S \ref{unSED}. Additionally, since our sample selection criterion of the AGN contribution is $f_{\rm AGN, IR}\geq 0.2$ (described in \S \ref{Sample}), we add extra choices of AGN contributions of 0.13, 0.15, and 0.18 during SED fitting to improve the accuracy of SED fittings near our threshold. The AGN contributions can take any of the following values during the SED fitting, from 0.0 to 0.99 as shown in Table \ref{tab:SED_config}.

For the dust attenuation, we follow the same parameter set in \cite{Wang2020} instead of the one in \cite{Yang2023}. \cite{Yang2023} reports that \textsc{CIGALE} may underestimate the redshift compared with spec-$z$ from CANDELS-EGS catalogue. However, to pursue an accurate estimation, we directly use spec-$z$ from the CANDELS-EGS catalogue if available ($\sim$23\% of our sample). For the remaining no-spec-$z$ sources, we give a range of redshift to allow \textsc{CIGALE} to estimate photo-$z$. 

CIGALE provides a Bayesian estimation for the output of physical properties. This estimation considers all possible models to weigh a statistically-based physical property, while the best-fit SED output is based on the minimal $\chi^2$ of the model. Statistically speaking, the Bayesian output is more reliable than the best fit. \cite{Yang2023} also highlights the robustness of the Bayesian output. Therefore, in this paper, we choose the Bayesian output for the analysis rather than the best fit.

\begin{figure*}
    \includegraphics[width=0.45\textwidth]{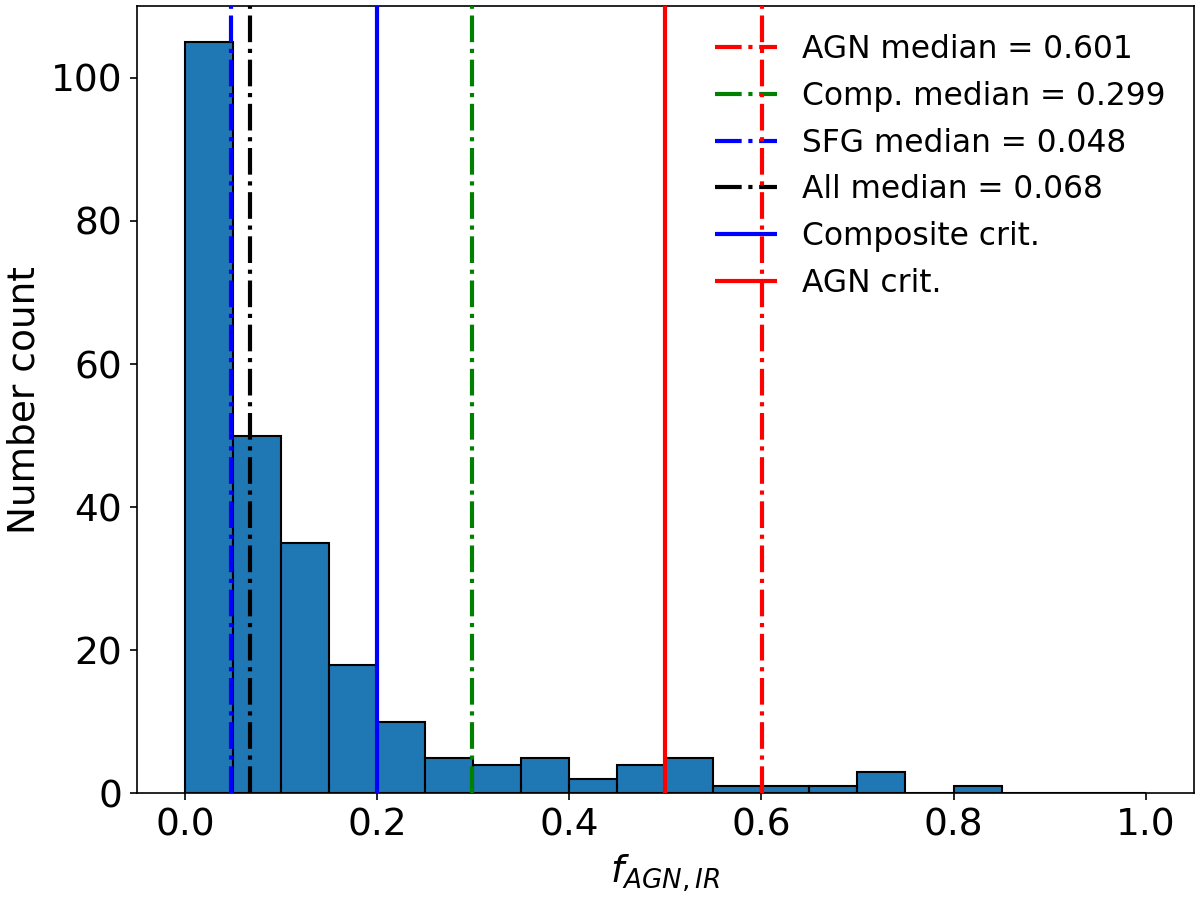}
    \includegraphics[width=0.45\textwidth]{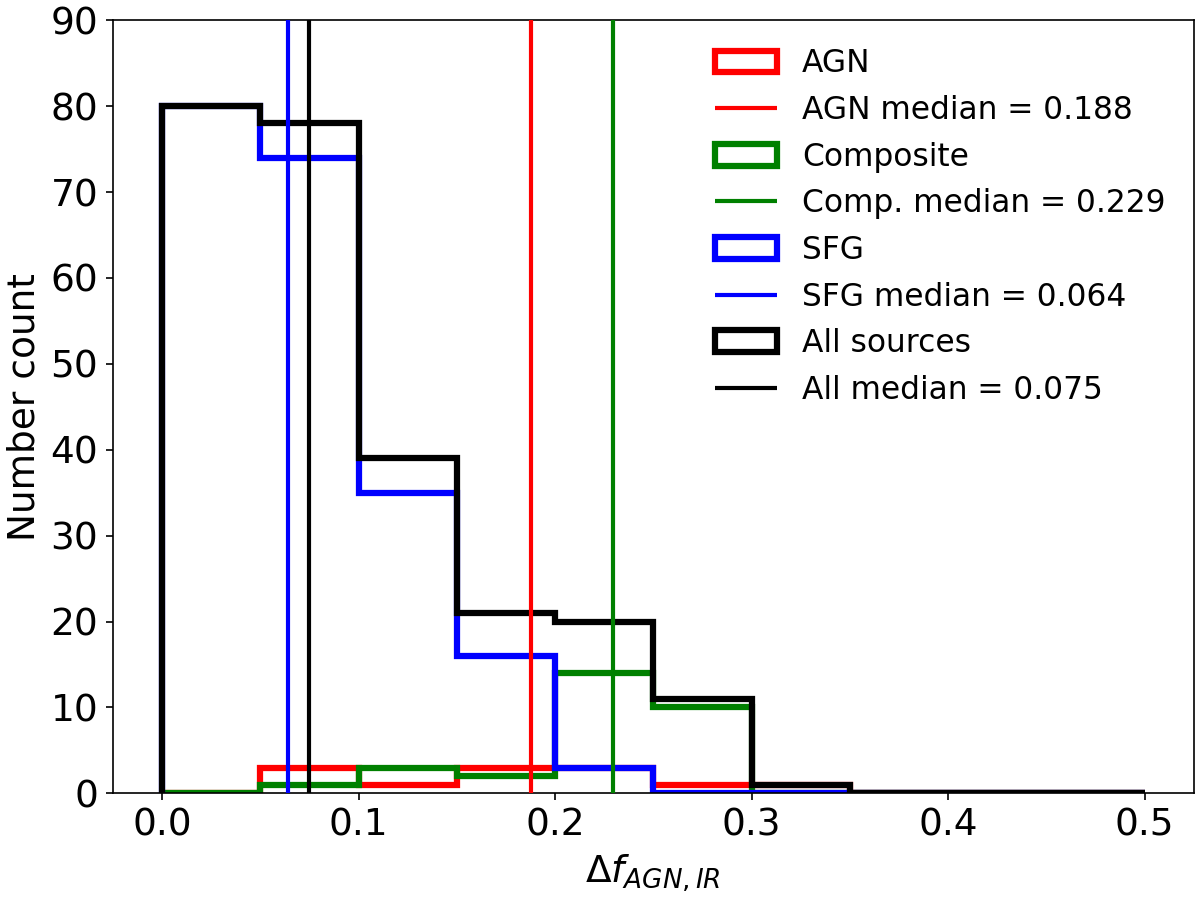}
    \caption{The histogram of $f_{\rm AGN, IR}$ and its error. The left panel shows the distribution of $f_{\rm AGN, IR}$. The solid lines mark the criteria for composite and AGN. The dot-dash lines show the median $f_{\rm AGN, IR}$ of AGN, composite, SFG, and all sources, respectively. The right panel shows the error of $f_{\rm AGN, IR}$ for AGN, composite, SFG, and all sources. The red, green, blue, and black solid lines are the median errors of $f_{\rm AGN, IR}$ for AGN, composite, SFG, and all sources, respectively.}
    \label{fig:fAGN_un}
\end{figure*}

According to the Bayesian $f_{\rm AGN, IR}$, we check the estimation quality. Fig.~\ref{fig:fAGN_un} displays the results. The median $f_{\rm AGN, IR}$ and error for all sources are $\sim$0.068 and $\sim$0.075, respectively. The error is not too large to affect our following analysis significantly as our selection criterion is $f_{\rm AGN, IR}\geq 0.2$, making our selection less likely to be severely contaminated by SFG.

Fig.~\ref{fig:SEDcut1} displays the performance of SED fittings with their MIRI cutouts.

\subsection{Sample selections}\label{Sample}
After our initial run of CIGALE on 573 sources we select our initial sample based on the SED fitting results.

First, our work mainly focuses on using mid-IR photometry to select AGNs, \cite{Yang2023} provides a criterion that each source should be detected by at least 2 bands in \textit{JWST} MIRI. However, to obtain a comprehensive AGN census and ensure that the MIR colour-colour diagram \citep[e.g.,][]{Kirkpartrick2017}{}{} is available, we impose a more stringent constraint, requiring at least 3-band detection.

Additionally, since two of the pointings (observation ID: o012\_t026 and o015\_t028) miss 7 $\mu m$ and 21 $\mu m$ detection, we are unable to apply the 7 $\mu m$ 80\% completeness flux limit, which has the faintest flux limit \citep{Ling2022, Wu2023}. Therefore, we adopt the second faintest 10 $\mu m$ 80\% completeness flux limit \citep{Ling2022, Wu2023} to exclude sources that fall below the flux limit. Some sources might have no 10 $\mu m$ detection by MIRI, so we substitute the estimated 10 $\mu m$ flux from \textsc{CIGALE} for undetected flux to make the completeness selection feasible and doable. 

\begin{figure}
    \includegraphics[width=\columnwidth]{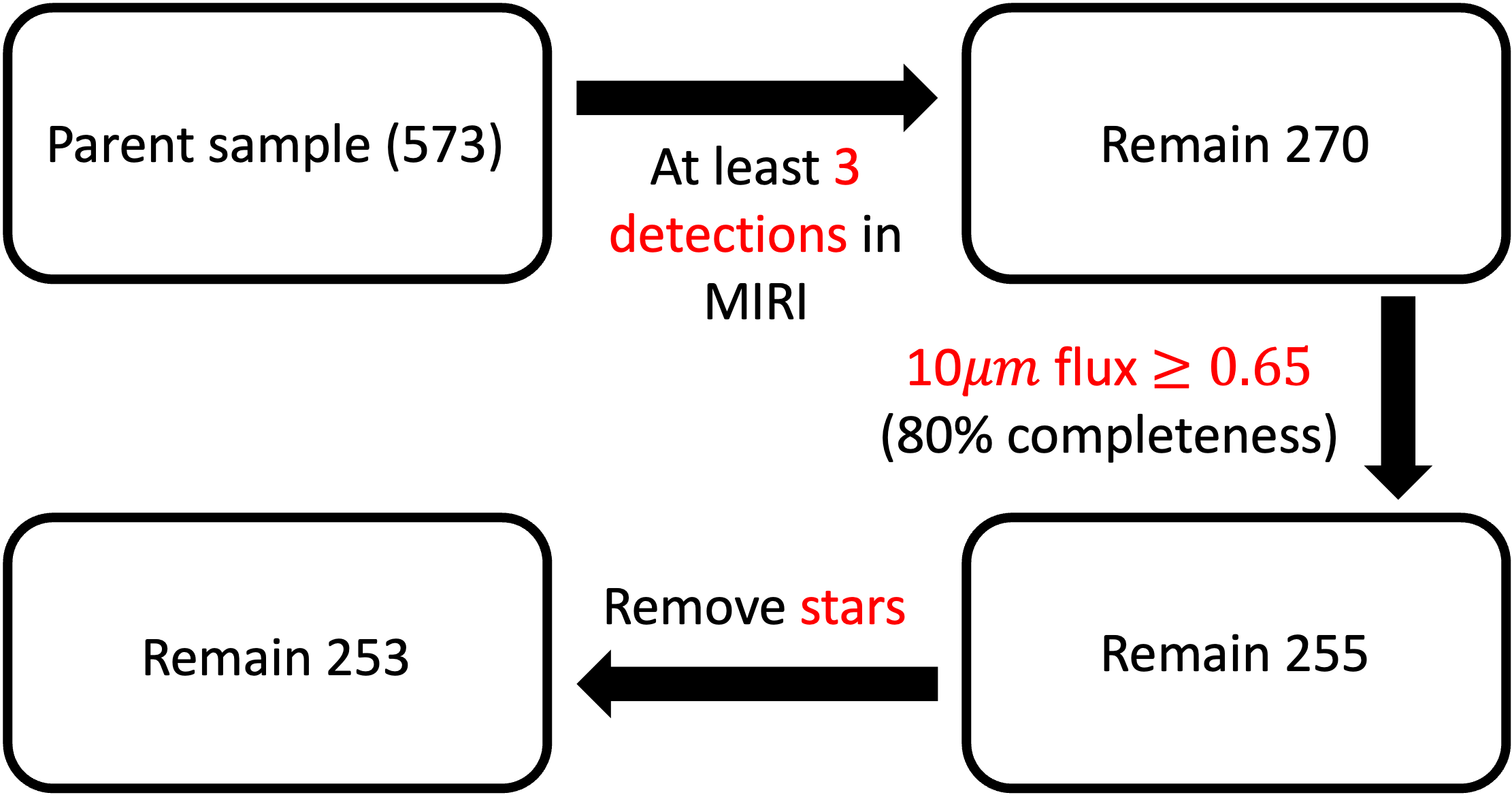}
    \caption{The flowchart of how we select our sample.}
    \label{fig:process}
\end{figure}

Finally, we check the performance of SED fittings and find out that there are 2 stars in the sample. In addition to stars, we find a source whose $\chi^2$ is 9.6, which is larger than the median $\chi^2$ of the rest of the sample ($\sim$1.14). After checking the image from the Hubble Space Telescope (\textit{HST}), Spitzer, and JWST, this source shows no abnormality and is included in our final sample. In summary, we exclude 2 stars. 253 sources remain in the final sample. Fig.~\ref{fig:process} shows the flowchart of our sample selections.

To define our sources as the candidates, we follow the same criterion as \cite{Wang2020}, requiring $f_{\rm AGN, IR}\geq0.2$ for the candidates. We choose this criterion to enable a proper comparison with \cite{Wang2020} and to demonstrate the unprecedented sensitivity of \textit{JWST} \citep[e.g.,][]{Kim2021, Ho2021}. Additionally, We divide our candidates into two based on the following criteria: composite ($0.2\leq f_{\rm AGN, IR} < 0.5$) and AGN ($f_{\rm AGN, IR}\geq 0.5$). Among 42 candidates, 30 of them are composites and 12 of them are AGNs. The remaining 210 sources are classified as SFGs because of low AGN contributions ($f_{\rm AGN, IR}< 0.2$).

\section{Results and discussion}\label{RandD}
\subsection{The uncertainties of SED fittings}\label{unSED}
For the discussion on $f_{\rm AGN, IR}$ in this work, the absence of FIR observations is a problem for SED fittings. The uncertainties may occur when the scientific results are analysed by physical properties estimated from SED fittings. To explore this influence, it is necessary to apply mock observations. In \textsc{CIGALE}, a mode called "savefluxes" can introduce a simulation to generate mock observations. We utilise the savefluxes model to generate 100 observations to analyse how significant the uncertainty is when the absence of FIR observations occurs. Table \ref{tab:SED_mock} lists the parameters applied to generate the mock observations. We set the majority of parameters to be one choice and merely change the AGN contribution and redshift since we focus on two crucial physical quantities in this work: AGN contribution and total IR luminosity to constrain the mock performance. Both quantities are strongly affected by the choice of AGN contribution and redshift parameters in \textsc{CIGALE}. Note that we set the redshift range from 0.5 to 1.0 rather than starting from 0.01 due to the avoidance of overflow in \textsc{CIGALE} if the redshift is smaller than 0.5.

\begin{figure}
    \includegraphics[width=\columnwidth]{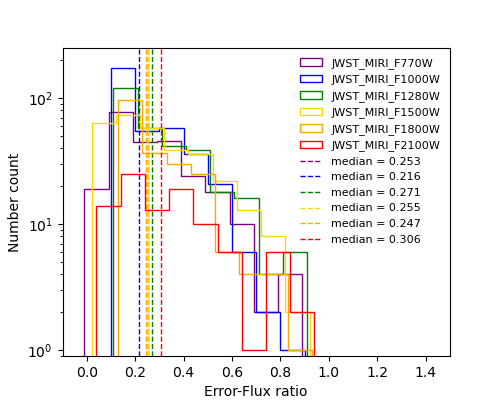}
    \caption{An example of the distributions of the error-flux ratio for \textit{JWST} MIRI. The dashed lines show the median for each band.}
    \label{fig:ratio_dis}
\end{figure}

We separate the mock observations into two groups. One group (hereafter, the group with-Herschel) contains the Spectral and Photometric Imaging Receiver \citep[SPIRE,][]{Griffin2010}{}{} PSW (250$\mu$m), PMW (350$\mu$m), and PLW (500$\mu$m) in the \textit{Herschel} Space Observatory (\textit{HSO}). The other group (hereafter, the group without-Herschel) does not. For the flux errors of the SPIRE mock observations, we adopt the $\rm SNR = 40$ flux-error relation from \cite{Pokhrel2016} to calculate the errors. For the other bands, we calculate the error-flux ratio for each band in each source. Fig.~\ref{fig:ratio_dis} shows the distribution of the error-flux ratio for \textit{JWST} as an example. We take the median value to generate the errors for each band.

\begin{figure*}
    \includegraphics[width=\columnwidth]{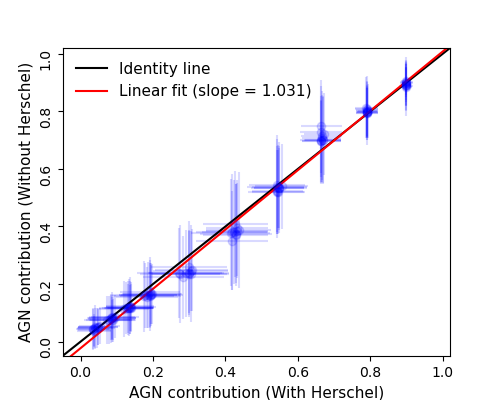}
    \includegraphics[width=\columnwidth]{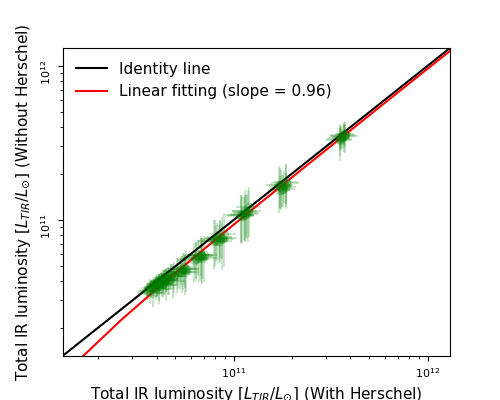}
    \caption{The comparison of both groups (with-Herschel and without-Herschel). The left panel shows the comparison of AGN contributions; The right panel shows the comparison of total IR luminosity. The solid black lines in the two panels represent the identity lines. The solid red line represents the linear fitting results.}
    \label{fig:two_comp}
\end{figure*}

\begin{figure*}
    \includegraphics[width=\columnwidth]{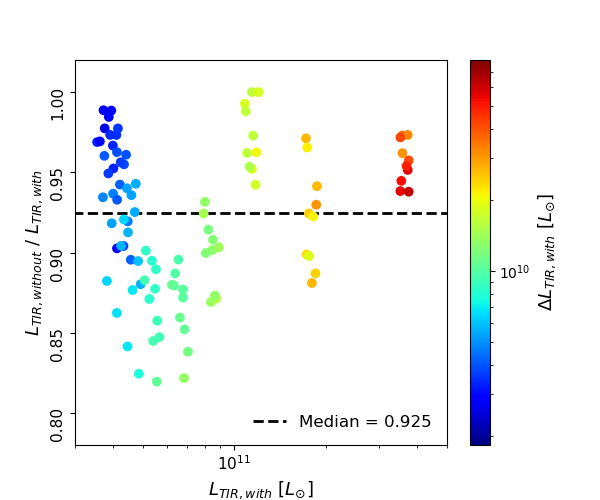}
    \includegraphics[width=\columnwidth]{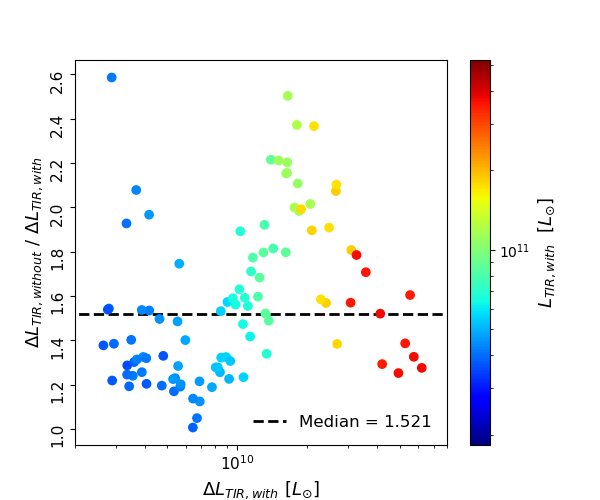}
    \includegraphics[width=\columnwidth]{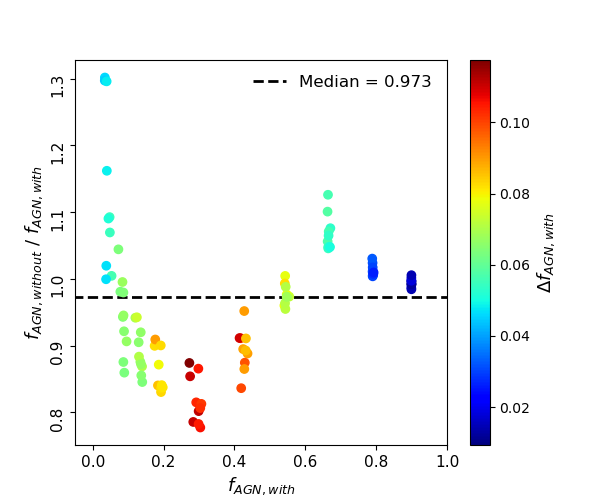}
    \includegraphics[width=\columnwidth]{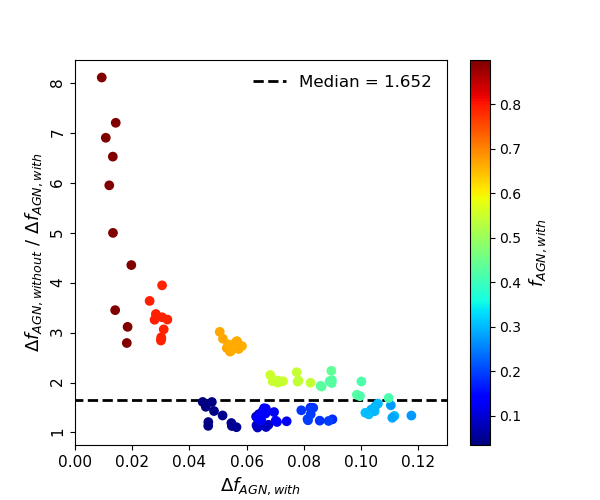}
    \caption{The detailed comparisons. The left panel compares total IR luminosity and AGN contribution; the right panel shows their error comparisons. The colour bar represents the corresponding errors or estimated values, respectively. The dashed black lines represent the median ratio for each block.}
    \label{fig:four_comp}
\end{figure*}

Next, we utilise the parameter list in Table \ref{tab:SED_config} to re-fit the SEDs for both groups and analyse their performance respectively. Fig.~\ref{fig:two_comp} shows their performance. Note that there are some small point displacements in both figures, this phenomenon is due to the choice of using Bayesian estimations for AGN contribution and total IR luminosity. Bayesian estimations can consider more situations than using input numbers. The estimated values for both groups are consistent according to the estimated values (for without-Herschel, $\sim$2.7\% smaller for AGN contribution and $\sim$7.5\% smaller for total IR luminosity) and the fitting slopes (1.031 for AGN contribution and 0.96 for total IR luminosity). However, for the error estimations, the median error of AGN contribution for the group with-Herschel (without-Herschel) is $\sim$0.065 ($\sim$0.104); The median error of total IR luminosity is $\sim9.7\times10^{9}$ ($\sim1.47\times10^{10}$) in solar luminosity $\rm L_{\odot}$. Fig.~\ref{fig:four_comp} shows detailed results, proving that the without-Herschel group estimates larger errors than the with-Herschel group ($\sim$65.2\% larger for AGN contribution and $\sim$52.1\% larger for total IR luminosity, both take the median value of the fraction in Fig.~\ref{fig:four_comp}). It indicates that the catalogue in which FIR observations are absent surely enlarges the uncertainty of estimated physical properties. In conclusion, although the uncertainty of errors is enlarged, we claim that lacking FIR observations in this work does not largely and significantly change our conclusions. Additionally, compared with the with-Herschel group, the estimated total IR luminosity in the without-Herschel group seems to be slightly underestimated ($\sim$7.5\%). The majority of points are below the ratio = 1.0. This underestimation might refer to the importance of FIR observations as well.


\begin{figure}
        \includegraphics[width=\columnwidth]{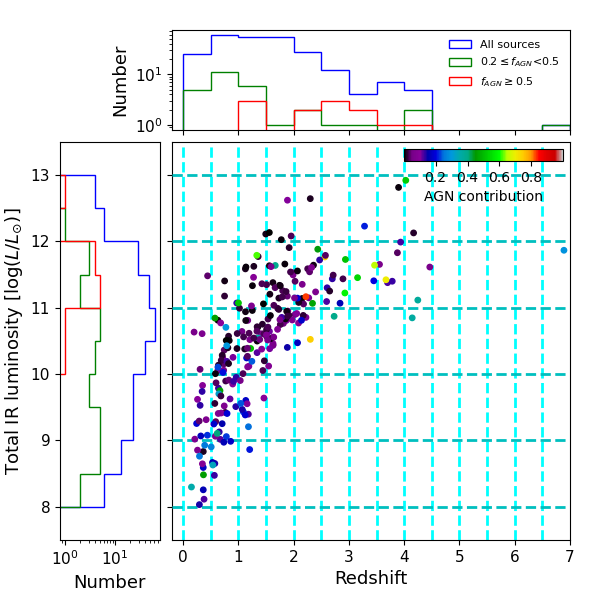}
    \caption{Total IR luminosity against redshift in this work. The histogram along the x-axis is the distribution of redshift. The histogram along the y-axis is the distribution of total IR luminosity in the logarithm. The colour bar shows the AGN contribution of each source. The cyan dashed lines cut the domain of redshift and total IR luminosity into several bins.}
    \label{fig:all_distribution}
\end{figure}
\subsection{Source distribution}\label{SD}
We start to discuss our fitting obtained through \textsc{CIGALE}. In the following subsections, we concentrate on two essential properties: AGN contribution and AGN number fraction. To establish the relations of AGN contribution and AGN number fraction with other physical properties, we have depicted a preliminary picture of our sources distribution in Fig.~\ref{fig:all_distribution}.  From Fig.~\ref{fig:all_distribution}, our AGN candidates are mostly located in higher redshift with higher AGN contribution rather than the composite candidates, which is consistent with the results in \cite{Toba2020}. However, it is not sufficient to claim this relation. An advanced analysis is necessarily introduced. To determine whether there are relations between AGN contribution and total IR luminosity more specifically, we investigate both AGN contribution and AGN number fraction in \S \ref{S:AGN_contri} and \S \ref{S:AGN_num}, respectively, to obtain a clearer understanding of their relations.

In addition, we compare our candidates with those of \cite{Yang2023}, who, before our work, also utilised the same field and sample to select AGN candidates. However, their criteria were $0.1\leq f_{\rm AGN, IR}<0.5$ for composite sources and $f_{\rm AGN, IR}\geq0.5$ for AGN. Applying these criteria to our candidates, the number of composites increases to 83, while the number of AGN remains at 12, yielding number ratios of $83/253 = 0.33\pm 0.04$ and $12/253 = 0.047\pm 0.014$, respectively. \cite{Yang2023} identified 102 composite sources and 25 AGN from 560 sources, resulting in $0.18\pm 0.02$ and $0.045\pm 0.009$ ratios for composites and AGN, respectively. While the AGN ratios are consistent, our composite ratio is higher.

The exact reason for this difference is unclear, but possible sources include the fact that \cite{Yang2023} used sources detected in at least two MIRI filters, whereas we required detections in three filters. This results in a smaller sample size for our analysis (253 vs. 560), which might lead to the selection of intrinsically brighter sources and a slightly larger fraction of composites. Furthermore, as shown in Fig.~\ref{fig:fAGN_un}, the number of objects changes drastically with small variations in $f_{\rm AGN, IR}$, which can occur due to slight differences in the fitting parameters between our analyses. This could also contribute to the observed difference in the fraction of composite sources.

\subsection{AGN contribution}\label{S:AGN_contri}
We calculate the average AGN contribution in each bin and check whether the relation between AGN contribution and redshift or total IR luminosity is clear or not. Note that we ignore the bins which have only one source for consistency and proper statistical meaning.

\begin{figure}
	\includegraphics[width=\columnwidth]{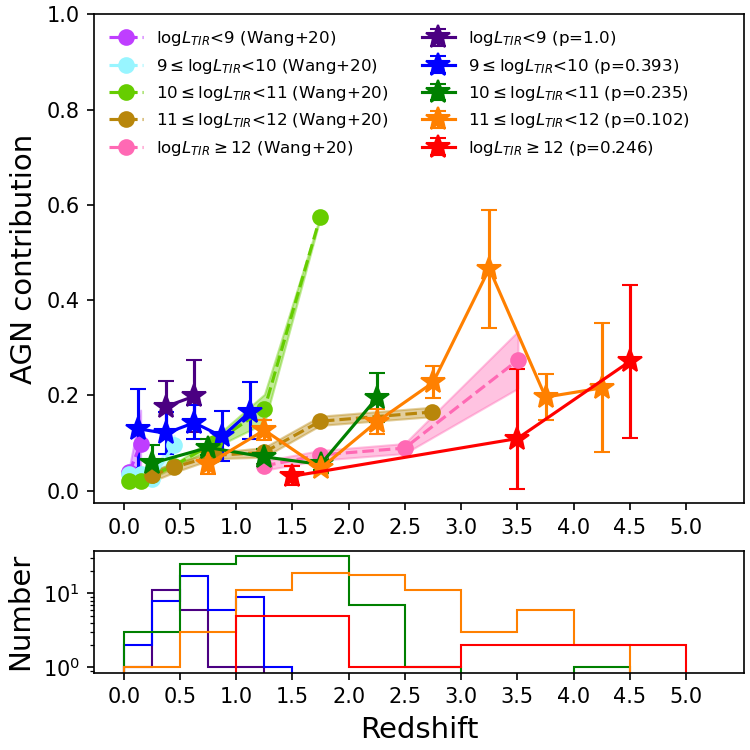}
    \caption{AGN contribution as a function of redshift. Circle markers are the results in \protect\cite{Wang2020}. Star markers are the results of this work. Purple is the $\log{L_{\rm TIR}}<9$ group. Blue is the $9\leq \log{L_{\rm TIR}}<10$ group. Green is the $10\leq \log{L_{\rm TIR}}<11$ group. Orange is the $11\leq \log{L_{\rm TIR}}<12$ group. Red is the $\log{L_{\rm TIR}}\geq12$ group. The bottom panel shows the histogram for each bin in each group. Here we also show the p-value of each group.}
    \label{fig:contri_redshift}
\end{figure}

Fig.~\ref{fig:contri_redshift} illustrates AGN contribution as a function of redshift. The results in \cite{Wang2020} are depicted as well to show the comparison. It seems that the green and the orange group are increasing. However, if the p-value is considered, both exceed the p-value threshold ($p\leq 0.05$), which means that we cannot conclude a tight relation between AGN contribution and redshift in this work statistically. This discrepancy with \cite{Wang2020} might be due to the lack of sources. Therefore, a larger sample size is required, enabling us to study the relationship more exhaustively.

\begin{figure}
	\includegraphics[width=\columnwidth]{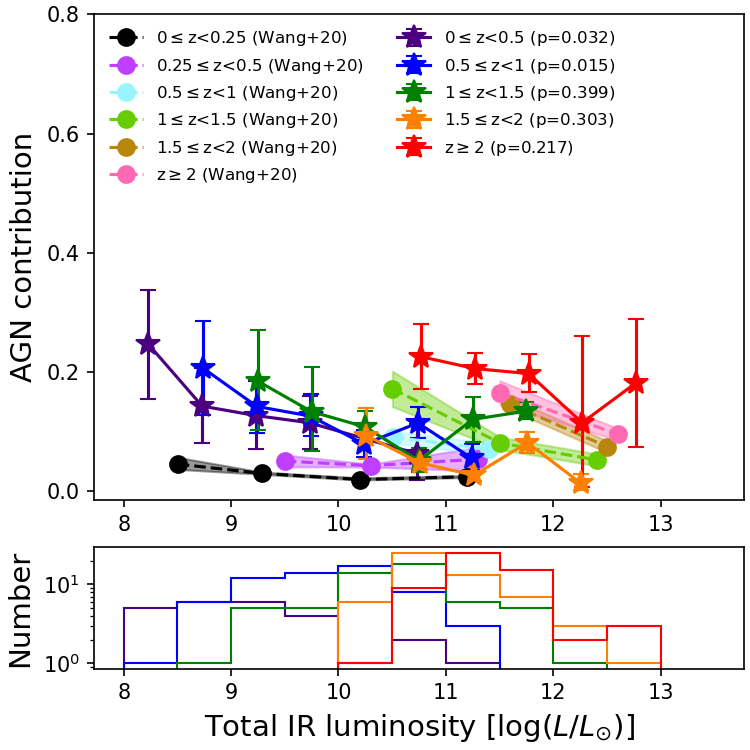}
    \caption{AGN contribution as a function of total IR luminosity. Circle markers are the results in \protect\cite{Wang2020}. Star markers are the results of this work. Purple is the $0\leq z<0.5$ group. Blue is $0.5\leq z<1$ group. Green is the $1\leq z<1.5$ group. Orange is the $1.5\leq z<2$ group. Red is the $z\geq2$ group. Black is the $0\leq z<0.25$ group in \protect\cite{Wang2020} which is merged in the purple group in this work. The bottom panel shows the histogram for each bin in each group. Here we also show the p-value for each group.}
    \label{fig:contri_TIR}
\end{figure}

Fig.~\ref{fig:contri_TIR} is AGN contribution as a function of total IR luminosity. It seems that AGN contribution is decreasing with increasing luminosity, especially for the purple and blue groups. This result is inconsistent with \cite{Chiang2019}. \cite{Chiang2019} reports that AGN contribution is increasing as the luminosity goes higher. Such difference might be due to the source distribution of the sample in this work. The majority of sources ($\sim$69.8\%) are below $10^{11} L_{\odot}$, which are not the typical luminous infrared galaxies \citep[LIRGs,][]{Sanders1996}{}{}. We claim that this reason might be the key to link to the decreasing result. Besides, all groups extend their tails to their faint ends compared with \cite{Wang2020}. This extension might be attributable to the level of sensitivity between \textit{AKARI} and \textit{JWST}. \textit{JWST} has a high sensitivity that can capture many fainter galaxies than before, inferring the power of finding fainter AGNs in the Universe. This achievement can allow us to investigate the evolution of AGNs and even the reionization of the Universe \citep{Giallongo2015, Grazian2018} further. On the other hand, Fig.~\ref{fig:contri_TIR} might indicate that \textit{JWST} has successfully accomplished our expectations.

\subsection{AGN number fraction}\label{S:AGN_num}
AGN number fraction ($f_{\rm num}$) can demonstrate how AGN populates in each redshift bin. The definition of AGN number fraction is presented in eq.~\ref{eq:eq2}:
\begin{equation}
    f_{\rm num} = \dfrac{N_{\rm Candidate}}{N_{\rm Candidate}+N_{\rm SFG}}.
    \label{eq:eq2}
\end{equation}
\begin{figure}
	\includegraphics[width=\columnwidth]{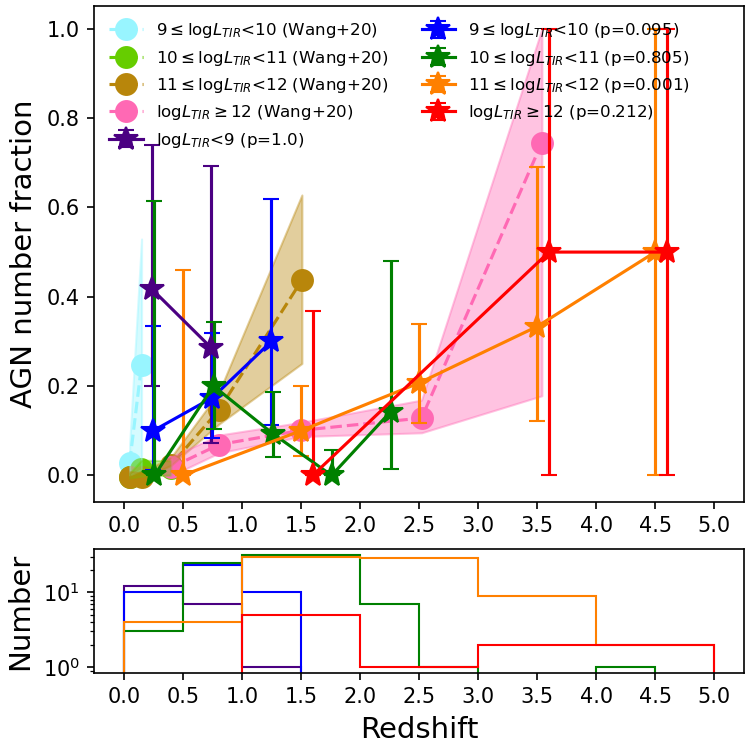}
    \caption{AGN number fraction as a function of redshift. Circle markers are the results in \protect\cite{Wang2020}. Star markers are the results of this work. Purple is the $\log{L_{\rm TIR}}<9$ group. Blue is the $9\leq \log{L_{\rm TIR}}<10$ group. Green is the $10\leq \log{L_{\rm TIR}}<11$ group. Orange is the $11\leq \log{L_{\rm TIR}}<12$ group. Red is the $\log{L_{\rm TIR}}\geq12$ group. The bottom panel shows the histogram for each bin in each group. Here we also show the p-value of each group.}
    \label{fig:num_redshoft}
\end{figure}
where $N_{\rm Candidate}$ is the number of our composite candidates plus AGN candidates. 

Fig.~\ref{fig:num_redshoft} shows our result. We compute the AGN number fraction in each bin and utilise the 1$\sigma$ errors for small statistics presented in \cite{Gehrels1986} to define the error bar for each bin. \cite{Wang2020} reports that AGN number fraction might have an increasing relation as a function of redshift. According to our results, only the orange group displays a low-p-value ($p\leq 0.05$) increasing relation. Nevertheless, the small sample size and small statistics make the uncertainty large enough to be unable to identify the truth of the trend. \cite{Kirkpatrick2023} reports that the AGN population is also increasing. However, their sample is also affected by the small statistics, making the uncertainty hard to constrain. In contrast, \cite{Hashiguchi2023} utilises much larger ($\sim27000$) sources to investigate the AGN number fraction in galaxy clusters. Their sample size is large enough to conclude a tightly increasing AGN number fraction against the redshift. To overcome the lack of sources, a larger mid-IR survey is necessary to increase more sources, lowering the difficulty of identifying scientific relations. However, the good news is that thanks to the high sensitivity of \textit{JWST}, we can unearth more high-$z$ galaxies that are difficult to observe by previous telescopes. Focusing on the blue, green, and orange groups, we conclude that we successfully extend and reveal the results in \cite{Wang2020} to the higher redshift. Besides, the result shows the difference between two different fields as well. 4 MIRI pointings ($\sim$ 8 arcmin$^2$) are much smaller than the NEP-wide field ($\sim$ 5.4 deg$^2$) in \cite{Wang2020}, the field of view limits the size of our sample in 4 MIRI pointings. This might be a key reason for our result. 
\begin{figure}
	\includegraphics[width=\columnwidth]{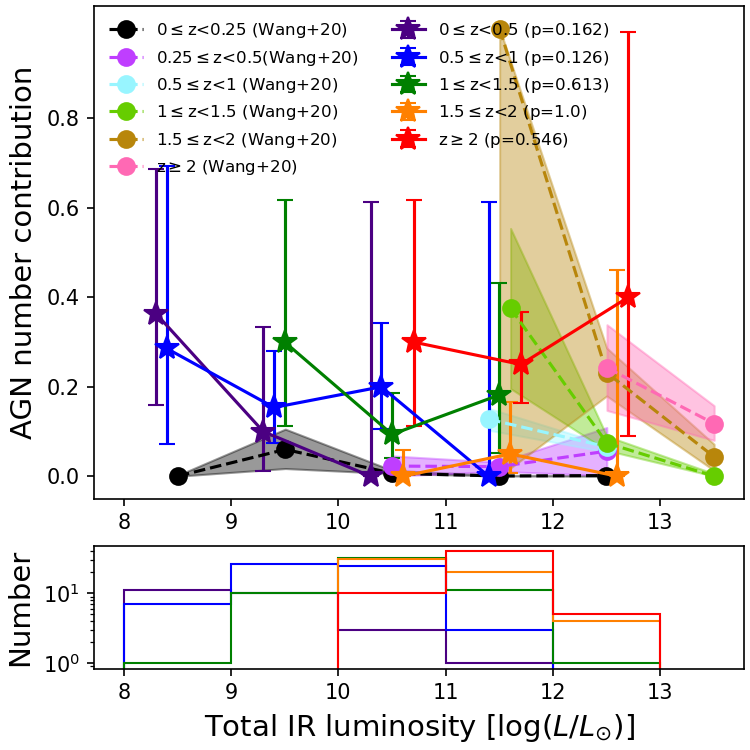}
    \caption{AGN number fraction as a function of total IR luminosity. Circle markers are the results in \protect\cite{Wang2020}. Star markers are the results of this work. Purple is the $0\leq z<0.5$ group. Blue is the $0.5\leq z<1$ group. Green is the $1\leq z<1.5$ group. Orange is the $1.5\leq z<2$ group. Red is the $z\geq2$ group. Black is the $0\leq z<0.25$ group in \protect\cite{Wang2020}, which is merged in the purple group in this work. Note that the lowest-z bin for this work is $0\leq z<0.5$ due to small statistics. The bottom panel shows the histogram for each bin in each group. We also show the p-value for each group.}
    \label{fig:num_TIR}
\end{figure}

In addition to redshift, we check AGN number fraction as a function of total IR luminosity shown in Fig.~\ref{fig:num_TIR}. Our result seems to have no obvious trends in different groups, which are consistent with \cite{Wang2020}. Furthermore, none of the p-values for each group are smaller than 0.05, leading us to difficulty identifying the existence of trends between AGN number fraction and total IR luminosity. However, if we focus on the groups except for the orange group, these groups show higher AGN number fractions at each fainter end. Nevertheless, \cite{Kartaltepe2010} utilises \textit{Spitzer} to investigate AGNs, reporting that the AGN number fraction is increasing at the ULIRGs and Hyper-luminous infrared galaxies \citep[HyLIRGs ($L_{\rm TIR}\geq 10^{13} L_{\odot}$), related work can be found in][]{Gao2021}{}{}. \cite{Chiang2019} concludes the same trend by \textit{AKARI} as well, reporting the luminous sources are dominated by the AGN population. These discrepancies might be due to the sensitivity difference between the previous telescope and \textit{JWST}, causing the previous works to be limited to low sensitivity. In this work, with the higher sensitivity of \textit{JWST}, we can extend our understanding down to fainter galaxies ($L_{\rm TIR}<10^{11} L_{\odot}$), not just LIRGs or ULIRGs. Moreover, as we mentioned above, the majority of the luminosity of our sources ($\sim$69.8\%) is not LIRGs. Consequently, combining the results in Fig.~\ref{fig:contri_TIR} and Fig.~\ref{fig:num_TIR}, they both suggest that we might successfully capture less luminous AGNs, which is a key goal (capturing fainter AGNs) in this work. This key result can lead us to believe that \textit{JWST} can unearth more faint AGNs in the distant Universe \citep[e.g.,][]{Harikane2023}{}{}. The majority of sources in \cite{Wang2020} are LIRG and ULIRG due to the limitation of sensitivity for \textit{AKARI}, and \cite{Kirkpatrick2023} reports the discovery of the population of mid-IR weak galaxies (where the mid-IR luminosity is mostly $\leq 10^{10} L_{\odot}$) in the high-z Universe. Combining with the results in this work, we can conclude that \textit{JWST} truly reveals a new era of faint AGN research.

\subsection{X-ray-selected AGNs}
In this section, we briefly discuss the X-ray-selected AGNs in our sample. An X-ray survey called AEGIS-X Deep (AEGIS-XD) survey by \textit{Chandra} X-ray observatory \citep{Nandra2015} provides a wild observation to investigate the properties of X-ray sources in the Universe, the CEERS field is within the survey area as well. 

Among our sample, AEGIS-XD has 13 matched X-ray sources. Adopting the X-ray AGN criterion by X-ray luminosity $L_{\rm X}\geq 10^{43}$ erg/s \citep{Auge2023}, we have 6 X-ray-selected AGN candidates and 4 of them are inside the candidates in this work ($\sim$66.6\%). The remaining 2 sources are classified as SFGs by our selection criteria, moreover, their AGN contributions are minimal ($f_{\rm AGN, IR}\leq 0.01$), which is quite lower than our criterion. Apart from the X-ray-selected AGNs, we still have 38 candidates not detected by the AEGIS-XD survey. However, with the MIRI on \textit{JWST}, we complement the gap of using X-ray to select AGNs.

\begin{figure}
	\includegraphics[width=\columnwidth]{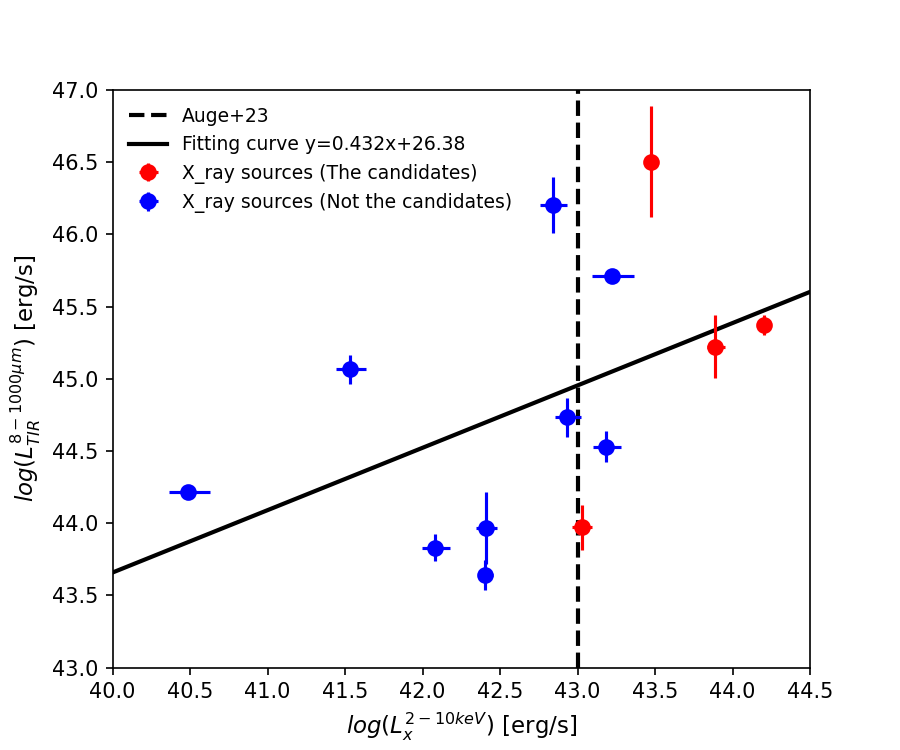}
    \caption{Relation between X-ray luminosity and total IR luminosity. The red dots are the candidates in this work. The blue dots are the SFGs in this work. The black dashed line shows the criterion of X-ray-selected AGN from \protect\cite{Auge2023}. The black solid line is the fitting curve.} 
    \label{fig:Lx&Ltir}
\end{figure}

Besides, we investigate the relation between X-ray luminosity and total IR luminosity. Fig.~\ref{fig:Lx&Ltir} illustrates the relation, a slightly increasing trend is clear. To confirm whether the relation is statistically reasonable, we apply the p-value test for X-ray and total IR luminosity. The result shows $\rm p = 0.928$, which is larger than 0.05, inferring that the relation in this work is statistically insignificant. 

\subsection{Colour-colour selections}
Here we adopt the \textit{JWST} MIRI colour-colour diagram for $z\sim 1$ ($0.75\leq z< 1.25$), $z\sim 1.5$ ($1.25\leq z< 1.75$), and $z\sim 2$ ($1.75\leq z< 2.25$) from \cite{Kirkpartrick2017} and compare to its performance. 

For the flux density ratio, we utilise the estimated flux value from \textsc{CIGALE} to replace the undetected flux for the feasible use of the colour-colour diagram. \cite{Kirkpartrick2017} provides different criteria to select these three objects: $f_{\rm AGN, MIR}<0.3$ as SFGs, $0.3\leq f_{\rm AGN, MIR}<0.7$ as composites, and $f_{\rm AGN, MIR}\geq 0.7$ as AGNs. 
Note that the aforementioned AGN contribution in \S \ref{CIGALE} is defined in $8-1000\mu m$ for total IR luminosity. \cite{Kirkpartrick2017} uses AGN mid-IR contribution ($f_{\rm AGN, MIR}$) which is defined in $5-15\mu m$ for mid IR luminosity. In Table \ref{tab:SED_config}, we do not particularly set the wavelength range of AGN contribution to be mid-IR range. Therefore, we integrate the mid-IR luminosity for dust emission and AGN emission from the best SED fitting result of our sample respectively, and calculate AGN mid-IR contribution using the same equation form as eq.\ref{eq:eq1}.

\begin{figure*}
	\includegraphics[width=\columnwidth]{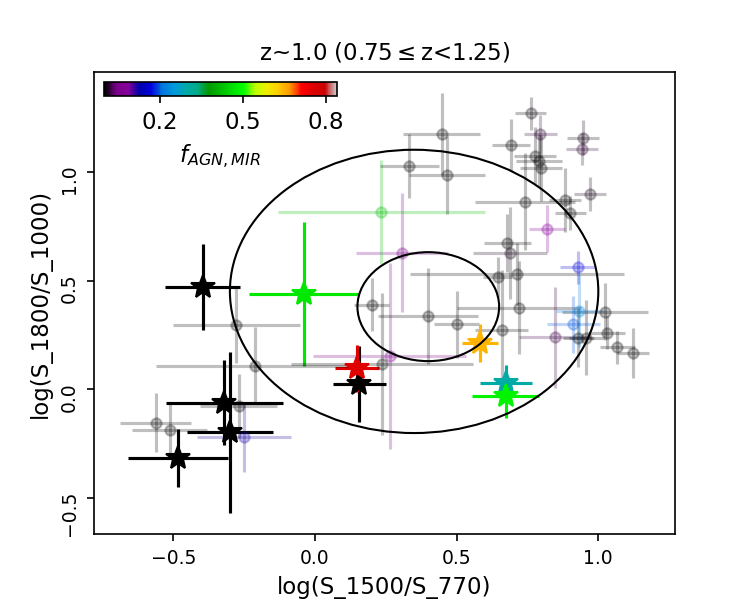}
        \includegraphics[width=\columnwidth]{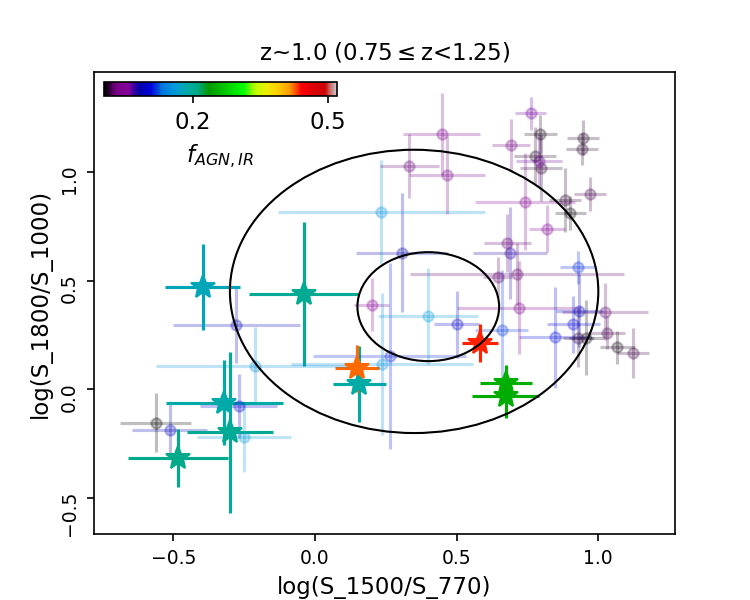}
        \caption{Colour-colour diagram at z$\sim$1 ($0.75\leq z<1.25$). The left panel shows the AGN mid-IR contribution ($f_{\rm AGN, MIR}$) for each source. The right panel shows the AGN contribution ($f_{\rm AGN, IR}$) for each source as well. Three regions divided by outer and inner circles present the SFGs, composites, and AGNs criterion from outside to inside defined in \protect\cite{Kirkpartrick2017}, respectively. The star markers are the candidates in this work.}
        \label{fig:z1_MIRcolour}
\end{figure*}

\begin{figure*}
	\includegraphics[width=\columnwidth]{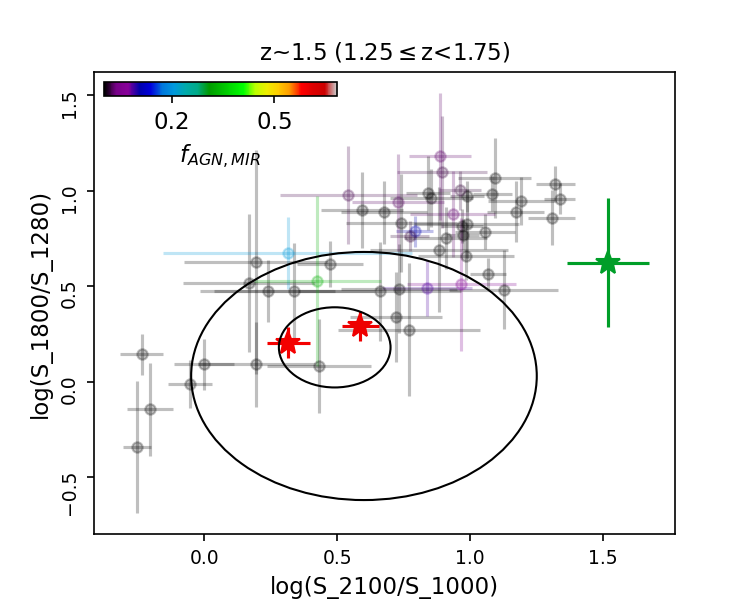}
        \includegraphics[width=\columnwidth]{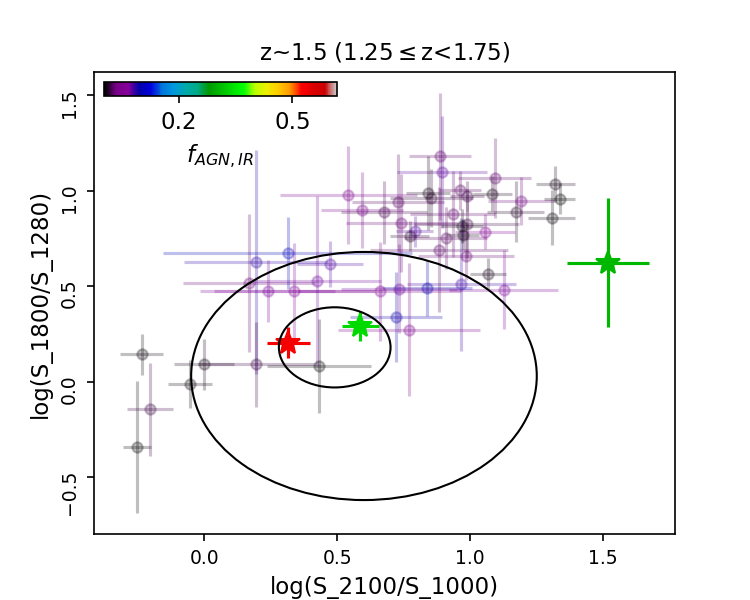}
        \caption{Colour-colour diagram at z$\sim$1.5 ($1.25\leq z<1.75$).}
        \label{fig:z15_MIRcolour}
\end{figure*}

\begin{figure*}
	\includegraphics[width=\columnwidth]{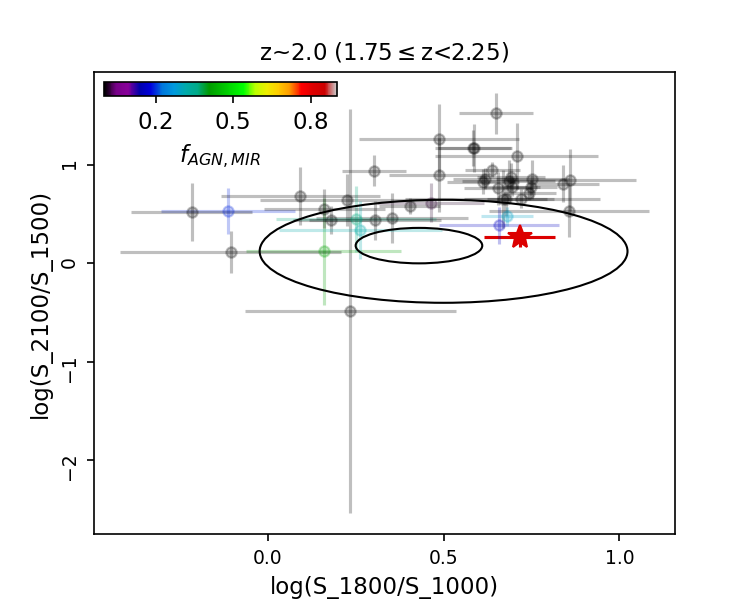}
        \includegraphics[width=\columnwidth]{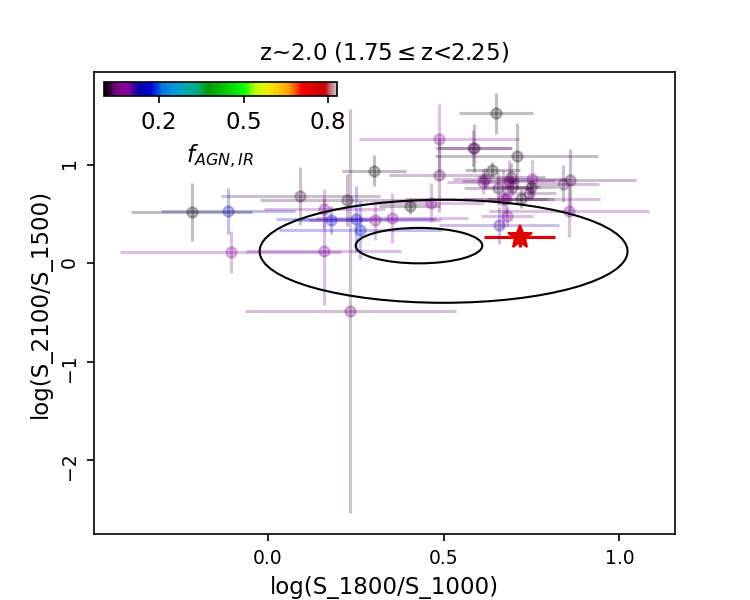}
        \caption{Colour-colour diagram at z$\sim$2 ($1.75\leq z<2.25$).}
        \label{fig:z2_MIRcolour}
\end{figure*}

\begin{figure*}
	\includegraphics[width=\columnwidth]{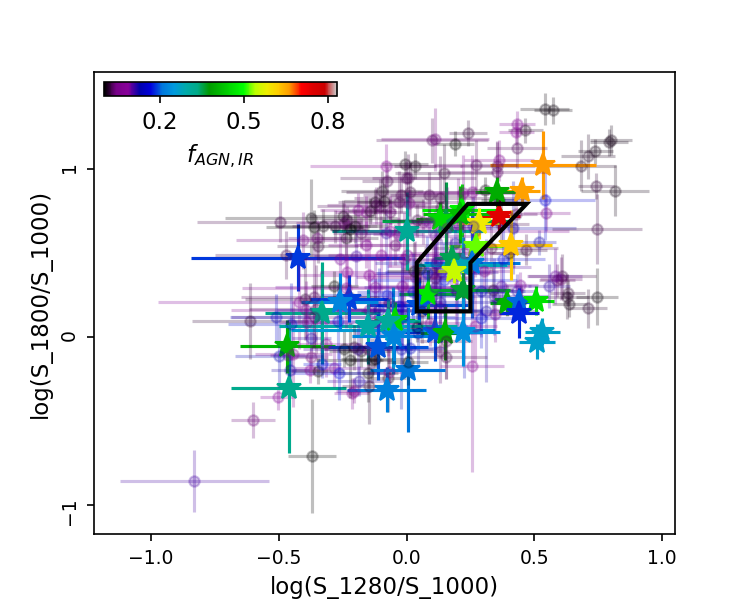}
        \includegraphics[width=\columnwidth]{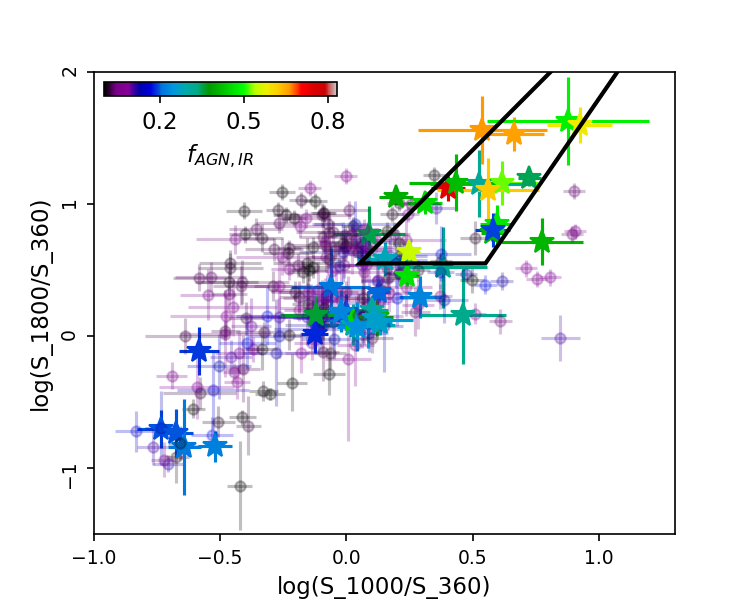}
        \caption{Two new colour-colour selections from \protect\cite{Kirkpatrick2023}. The left panel shows the new mid-IR selection which is not limited by the redshift; The right panel shows the near-IR + mid-IR combination. The near-IR band is IRAC1 in \textit{Spitzer} \protect\citep{Fazio2004}. The star markers are the candidates in this work. The black polygons display both AGN criteria defined in \protect\cite{Kirkpatrick2023}, respectively.} 
        \label{fig:new_MIR_NIR_colour}
\end{figure*}

Fig.~\ref{fig:z1_MIRcolour}, Fig.~\ref{fig:z15_MIRcolour}, and Fig.~\ref{fig:z2_MIRcolour} illustrate the results of our sample for both AGN mid-IR contribution and AGN contribution. The majority of our candidates are successfully captured by the colour-colour diagram. However, \cite{Kirkpatrick2023} recently reported that this selection can be contaminated by SFGs and mid-IR weak galaxies which can pretend to be the AGN candidates or have unknown redshift. Therefore, \cite{Kirkpatrick2023} provides two new colour-colour selections (hereafter, MIR-only and NIR+MIR, respectively) to avoid these influences. We apply both methods to select the AGN candidates (hereafter, the cc-AGNs) and compare them with our candidates (hereafter, the SED-AGNs). Fig.~\ref{fig:new_MIR_NIR_colour} shows both results, having 32 (32) cc-AGNs for MIR-only (NIR+MIR) selection. The number of overlapped candidates with the SED-AGNs is 10 (15), showing some missed candidates in both the cc-AGNs and SED-AGNs. We check the missed candidates the SED-AGNs lose and the cc-AGNs lose. The SED-AGNs miss 22 (17) candidates and the cc-AGNs miss 32 (27) candidates.

Given the missed candidates, we check whether they are SFGs for each set of candidates. We focus on possibly the most problematic cases, six SED-AGNs that both cc-selections classified as SFGs. We crosscheck the SED results and the AGN contributions for the 6 candidates. These 6 SED-AGN candidates missed by the cc-selections exhibit small AGN contributions close to our selection criterion ($f_{\rm AGN, IR}\sim 0.2$). Table \ref{tab:missSFGs} displays the detailed values and the relative errors. The errors are also large, while the best-fit AGN contributions for each source are also small at almost 0.0. As mentioned in \S \ref{CIGALE}, Bayesian $f_{\rm AGN, IR}$ is considered more robust than the best-fit $f_{\rm AGN, IR}$ statistically. With this consideration, the cc- and SED-AGNs are not overly inconsistent.

According to the above results, the cc-AGNs lose more candidates than the SED-AGNs, especially in the MIR selection. \cite{Kirkpatrick2023} mentions that the reliability of the NIR+MIR selection is higher than the MIR-only selection, which is consistent with our results. Additionally, \cite{Kirkpatrick2023} points out that applying the colour-colour selection in the small survey is not an efficient way to find AGNs, which agrees with our result of using MIR colour and NIR+MIR colour selection. Therefore, a larger survey \citep[e.g., COSMOS-Web,][]{Casey2023}{}{} is more suitable and might be a potentially useful data to apply colour-colour selections.

\begin{table*}
	\centering
	\begin{tabular}{ccc}
		\hline
		Source ID & Bayesian $f_{\rm AGN, IR}$ & Best-fit $f_{\rm AGN, IR}$ \\
		\hline
		CANDELS\_EGS\_F160W\_J141919.9+524851.5 & $0.25\pm 0.23$ & 0.0 \\
        CANDELS\_EGS\_F160W\_J141921.6+524839.1 & $0.21\pm 0.23$ & 0.0 \\
        CANDELS\_EGS\_F160W\_J142039.6+530330.7 & $0.21\pm 0.21$ & 0.0 \\
        CANDELS\_EGS\_F160W\_J141919.2+524858.7 & $0.24\pm 0.25$ & 0.0 \\
        CANDELS\_EGS\_F160W\_J141904.7+524853.8 & $0.22\pm 0.21$ & 0.0 \\
        CANDELS\_EGS\_F160W\_J141900.6+525002.9 & $0.22\pm 0.21$ & 0.0 \\
        \hline
	\end{tabular}
        \caption{The detailed information of the candidates that the cc-AGNs lose and are classified as the AGN candidates in this work.}
        \label{tab:missSFGs}
\end{table*}


\section{Conclusions}\label{conclusion}

We utilised a multi-wavelength merged catalogue in the CEERS survey and performed \textsc{CIGALE} SED fittings to investigate the physical properties of our sample (253 sources after the selections). Applying our criteria, we found 42 candidates and separated them into two categories according to $f_{\rm AGN, IR}$: 30 composites ($0.2\leq f_{\rm AGN, IR}<0.5$) and 12 AGNs ($f_{\rm AGN, IR}\geq 0.5$). We find previously reported relations in \cite{Wang2020} are not statistically obvious in this work due to their small sample size. This inconsistency might be attributed to the high sensitivity of JWST. JWST allows us to discover more fainter sources than previous works \citep{Wang2020, Chiang2019, Kartaltepe2010}.

In addition, we apply the MIRI colour-colour diagrams from \cite{Kirkpartrick2017} and \cite{Kirkpatrick2023}, where the majority of our candidates located in $0.75\leq z\leq 2.25$ are successfully captured by these diagrams. Recently, \cite{Kirkpatrick2023} introduced two more reliable colour-colour selections. They reported that the previous selection in \cite{Kirkpartrick2017} is unreliable and might be potentially affected by the contamination from SFGs and mid-IR weak galaxies. Moreover, it is also limited in certain redshift ranges, which is not general enough in the high-z Universe. Our findings reveal that the candidates with low AGN contribution ($f_{\rm AGN, IR}<0.5$, which is the composite candidates) failed to be identified as AGNs through colour-colour selections. This failure suggests that such methods might not be suitable for a small survey like CEERS in this work, which is consistent with the conclusion from \cite{Kirkpatrick2023}.

Note that the sample size in this work is not large enough ($\sim$ 573 sources), e.g., the relation in Fig.~\ref{fig:num_TIR} is still not obvious. Therefore, a larger sample size is required to confirm the relations investigated in this work. With more \textit{JWST} MIRI observations to be released, we expect that we can explore the properties of AGNs more accurately in the future.

\section*{Acknowledgements}
The authors express their appreciation to the anonymous referee for the constructive comments and suggestions which significantly improved the quality of the paper.

TG acknowledges the support of the National Science and Technology Council of Taiwan through grants 108-2628-M-007-004-MY3, 110-2112-M-005-013-MY3, 112-2112-M-007-013, and 112-2123-M-001-004-. 
TH acknowledges the support of the National Science and Technology Council of Taiwan through grants 110-2112-M-005-013-MY3, 110-2112-M-007-034-, 111-2123-M-001-008-, 110-2112-M-005-013-MY3, 110-2112-M-007-034-, and 112-2123-M-001-004-. 
SH acknowledges the support of The Australian Research Council Centre of Excellence for Gravitational Wave Discovery (OzGrav) and the Australian Research Council Centre of Excellence for All Sky Astrophysics in 3 Dimensions (ASTRO 3D), through project number CE17010000 and CE170100013, respectively.

This work is based on observations made with the NASA/ESA/CSA James Webb Space Telescope. The data were obtained from the Mikulski Archive for Space Telescopes at the Space Telescope Science Institute, which is operated by the Association of Universities for Research in Astronomy, Inc., under NASA contract NAS 5-03127 for JWST. These observations are associated with program JWST-ERS01345. This work is based on observations taken by the CANDELS Multi-Cycle Treasury Program with the NASA/ESA HST, which is operated by the Association of Universities for Research in Astronomy, Inc., under NASA contract NAS 5-26555.

This work used high-performance computing facilities operated by the Centre for Informatics and Computation in Astronomy (CICA) at National Tsing Hua University. This equipment was funded by the Ministry of Education of Taiwan, the National Science and Technology Council of Taiwan, and the National Tsing Hua University.

\section*{Data Availability}
The MIRI observations from \textit{JWST} CEERS survey are publicly available at the MAST archive \url{https://mast.stsci.edu/portal/Mashup/Clients/Mast/Portal.html}. The EGS Multi-Band Source and Photometric Redshift catalogue can be downloaded at \url{https://archive.stsci.edu/hlsp/candels/egs-catalogs}. Other data utilised in this work can be provided by request to the authors.



\bibliographystyle{mnras}
\bibliography{mnras_template} 




\appendix
\section{The cutouts of AGN candidates}
\begin{figure*}
	\includegraphics[width=0.85\textwidth]{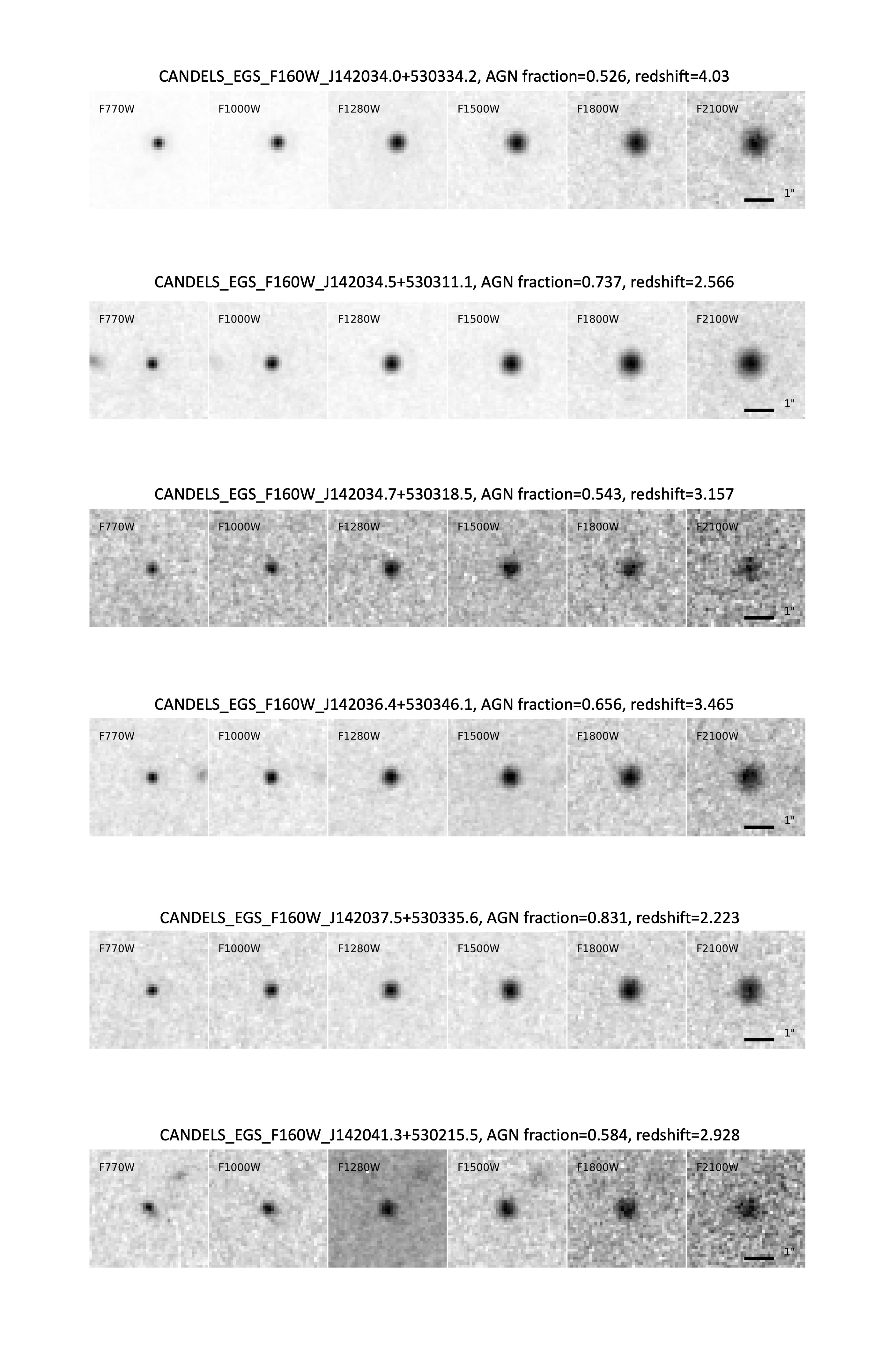}
        \caption{The AGN cutouts in the CEERS field pointings 1 (observation ID: o001\_t021).} 
        \label{fig:p1}
\end{figure*}

\begin{figure*}
	\includegraphics[width=0.85\textwidth]{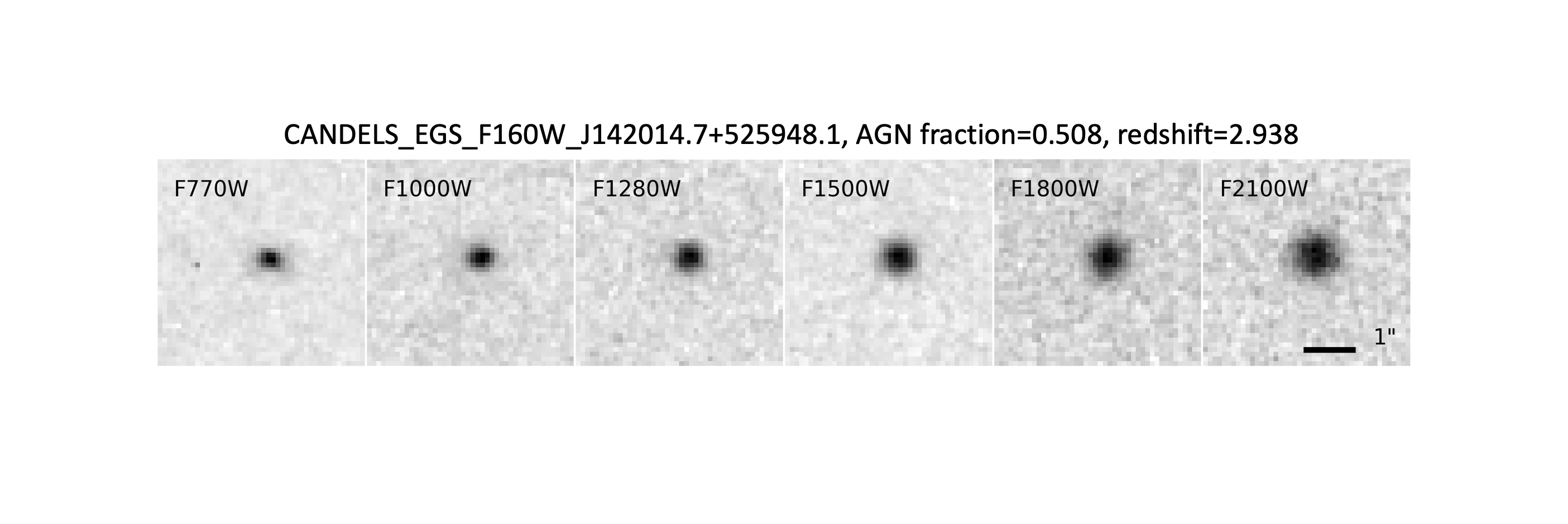}
        \caption{The AGN cutout in the CEERS field pointings 2 (observation ID: o002\_t022).} 
        \label{fig:p2}
\end{figure*}

\begin{figure*}
	\includegraphics[width=0.85\textwidth]{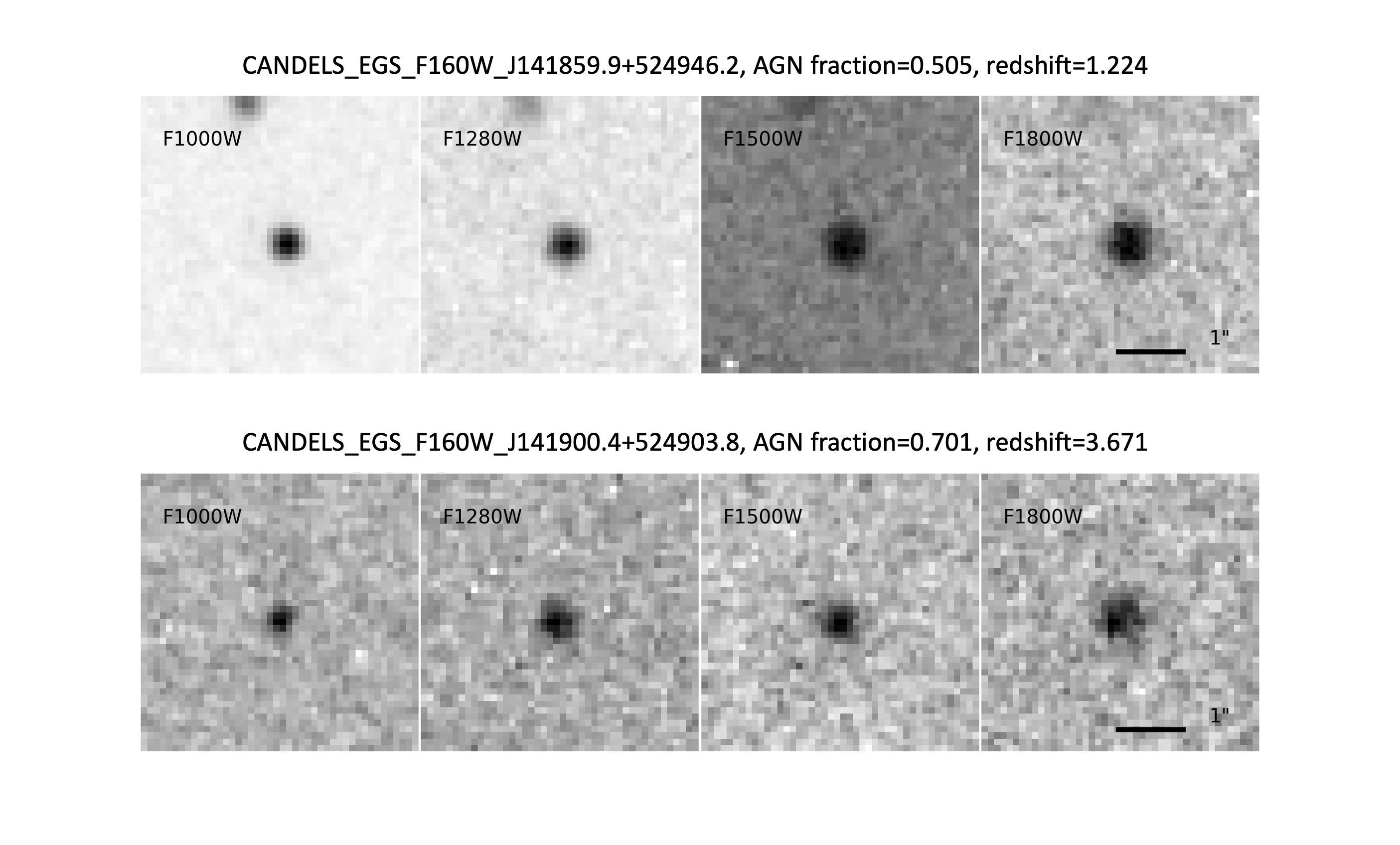}
        \caption{The AGN cutouts in the CEERS field pointings 5 (observation ID: o012\_t026).} 
        \label{fig:p5}
\end{figure*}

\begin{figure*}
	\includegraphics[width=0.85\textwidth]{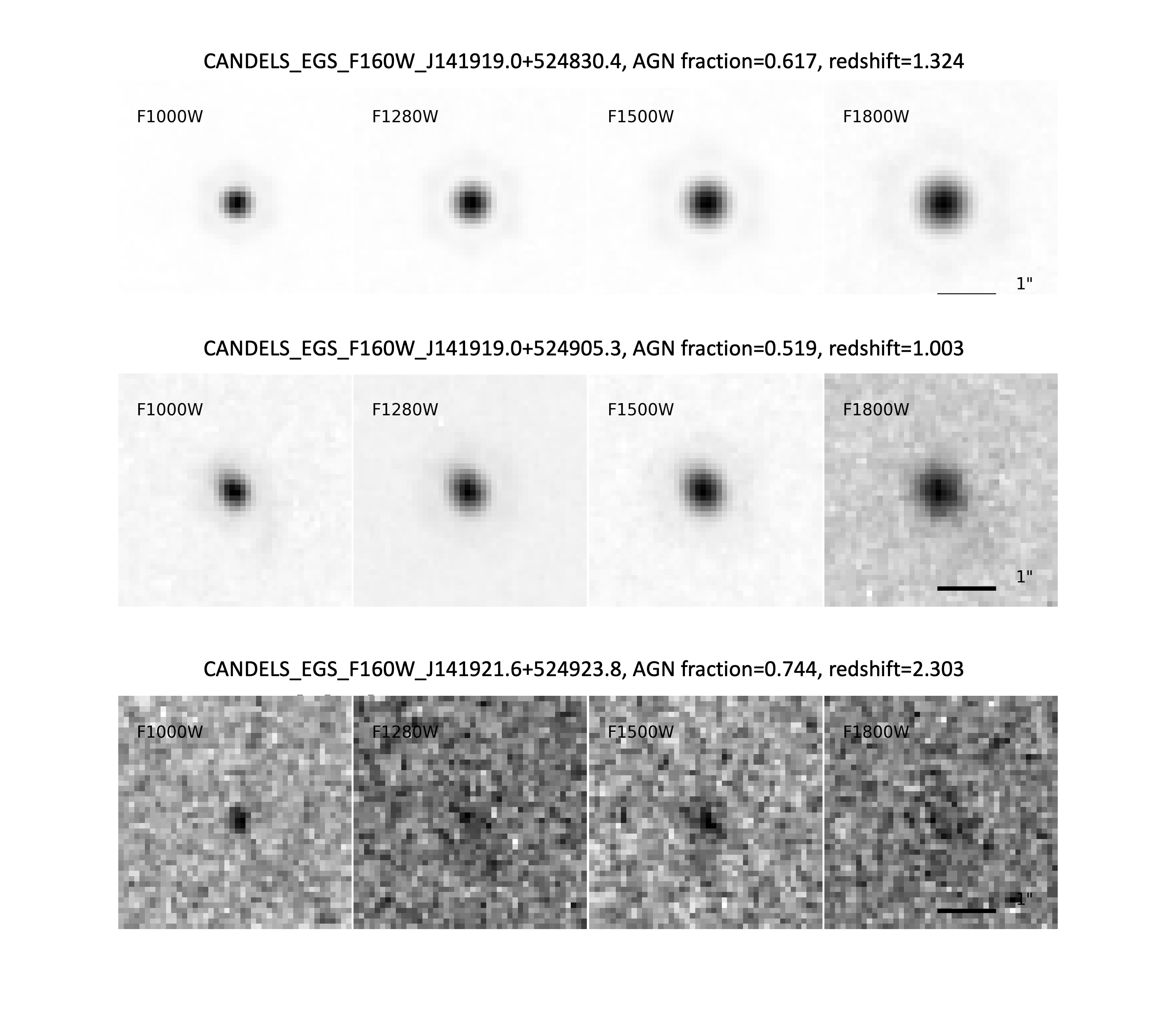}
        \caption{The AGN cutouts in the CEERS field pointings 8 (observation ID: o015\_t028).} 
        \label{fig:p8}
\end{figure*}

Here we show the cutouts of the AGN candidates ($f_{\rm AGN, IR}\geq 0.5$) by each pointings in Fig.~\ref{fig:p1}, Fig.~\ref{fig:p2}, Fig.~\ref{fig:p5}, and Fig.~\ref{fig:p8}.

\section{\textsc{CIGALE} parameter list}
In this appendix section, we list out the parameters used in SED fitting and generating mock observations in Table \ref{tab:SED_config} and Table \ref{tab:SED_mock}.
\begin{table*}
	\centering
	\begin{tabular}[c]{lllll}
		\hline
		Module & & Parameter & & Values\\
		\hline
		\texttt{sfhdelayed} & & $\tau_{\rm main}$ (Myr)  & & 100.0, 1000.0, 2000.0, 3000.0, 4000.0, 5000.0\\
	    ...                 & & age$_{\rm main}$ (Myr)   & & 100.0, 200.0, 500.0, 1000.0, 2000.0, 3000.0, 4000.0, 5000.0\\
            ...                 & & $\tau_{\rm burst}$ (Myr) & & 50.0\\
            ...                 & & age$_{\rm burst}$ (Myr)  & & 20.0\\
            ...                 & & f$_{\rm burst}$          & & 0.0\\
            ...                 & & sfr$_{\rm A}$            & & 1.0\\
            \hline
            \texttt{bc03}       & & Initial mass function & & \cite{1955ApJ...121..161S}\\
            ...                 & & metallicity           & & 0.02\\
            ...                 & & separation\_age       & & 10\\
            \hline
            \texttt{dustatt\_modified\_CF00} & & $A_{\rm v}$\_ISM (log scale) & & -2.0 to 1.0 (10 steps)\\
            ...                            & & $\mu$             & & 0.44\\
            ...                            & & slope\_ISM        & & -0.5, -0.7, -0.9\\
            ...                            & & slope\_BC         & & -0.7, -1.0, -1.3\\
            \hline
            \texttt{nebular}  & & logU            & & -2.0\\
            ...               & & Gas metallicity & & 0.02\\
            \hline
            \texttt{dl2014} & & q$_{\rm PAH}$ & & 0.47, 2.50, 7.32\\
            ...             & & U$_{\rm min}$ & & 0.1, 1.0, 10.0, 50.0\\
            ...             & & $\alpha$  & & 2.0\\
            ...             & & $\gamma$  & & 0.01, 0.02, 0.05, 0.1, 0.2, 0.5, 0.9\\
            \hline 
            \texttt{skirtor2016} & & Optical depth at 9.7$\mu$m ($\tau$) & & 3, 5, 7, 9, 11\\
            ...                  & & Opening angle & & 40\\
            ...                  & & Inclination   & & 70\\
            ...                  & & $f_{\rm AGN, IR}$  & & 0.0, 0.01, 0.05, 0.1, 0.13, 0.15, 0.18, 0.2\\
            ...                  & &  & & 0.3, 0.4, 0.5, 0.6, 0.7, 0.8, 0.9, 0.99\\
            \hline
            \texttt{redshift} & & redshift & & 0.01-7.0 (100 steps) for no-spec-$z$ sources\\ 
            \hline
	\end{tabular}
        \caption{Model parameters, note that parameters which are not listed here remain default.}
        \label{tab:SED_config}
\end{table*}

\begin{table*}
	\centering
	\begin{tabular}[c]{lll}
		\hline
		Module & Parameter & Values\\
		\hline
		\texttt{sfhdelayed} & $\tau_{\rm main}$ (Myr) & 500.0\\
	    ...                 & age$_{\rm main}$ (Myr)  & 4000.0\\
            \hline
            \texttt{dustatt\_modified\_CF00} & $A_{\rm v}$\_ISM (log scale) & 1.0\\
            ...                            & slope\_ISM        & -0.5\\
            ...                            & slope\_BC         & -1.3\\
            \hline
            \texttt{dl2014} & q$_{\rm PAH}$ & 7.32\\
            ...             & U$_{\rm min}$ & 0.1\\
            ...             & $\alpha$  & 2.0\\
            ...             & $\gamma$  & 0.01\\
            \hline 
            \texttt{skirtor2016} & Optical depth at 9.7$\mu$m ($\tau$) & 11\\
            ...                  & $f_{\rm AGN, IR}$  & 0.0-0.9 (10 steps)\\
            \hline
            \texttt{redshift} & redshift & 0.5-1.0 (10 steps)\\ 
            \hline
	\end{tabular}
        \caption{Model parameters used to generate mock observations. Parameters that are not listed here remain default and identical to Table \ref{tab:SED_config}.}
        \label{tab:SED_mock}
\end{table*}




\bsp	
\label{lastpage}
\end{document}